\documentclass[preprintnumbers, floatfix,letterpaper,aps,prd,epsfig,nofootinbib,longbibliography,twocolumn]{revtex4-1}

\usepackage{bm,graphicx,dcolumn,epstopdf,epsf, latexsym,mathbbol, amssymb,amsmath,color,slashed,mathrsfs,mathcomp,simplewick,caption}
\usepackage[center]{subfigure}
\usepackage{multirow}
\usepackage{makecell}
\usepackage{booktabs}
\usepackage{float}

\usepackage[colorlinks,linkcolor=blue,citecolor=blue,urlcolor=blue]{hyperref}

\begin{document}
\allowdisplaybreaks

\newlength\scratchlength
\newcommand\s[2]{
  \settoheight\scratchlength{\mathstrut}%
  \scratchlength=\number\numexpr\number#1-1\relax\scratchlength
  \lower.5\scratchlength\hbox{\scalebox{1}[#1]{$#2$}}%
}

\title{Equatorial periodic orbits and gravitational waveforms in a black hole free of Cauchy horizon}

\author{Chao-Hui Wang$^{a,b,c}$}

\author{Xiang-Cheng Meng$^{a,b,c}$}
\author{Yu-Peng Zhang$^{a,b,c}$}

\author{Tao Zhu$^{d,e}$}

\author{Shao-Wen Wei$^{a,b,c}$}
\email{weishw@lzu.edu.cn, corresponding author}

\affiliation{
$^{a}$Lanzhou Center for Theoretical Physics, Key Laboratory of Theoretical Physics of Gansu Province, School of Physical Science and Technology, Lanzhou University, Lanzhou 730000, People's Republic of China.\\
$^{b}$Institute of Theoretical Physics $\&$ Research Center of Gravitation,Lanzhou University, Lanzhou 730000, People's Republic of China.\\
$^{c}$ Gansu Provincial Research Center for Basic Disciplines of Quantum Physics.\\
$^{d}$Institute for Theoretical Physics and Cosmology, Zhejiang University of Technology, Hangzhou, 310023, People's Republic of China.\\
$^{e}$United Center for Gravitational Wave Physics (UCGWP), Zhejiang University of Technology, Hangzhou, 310023, People's Republic of China.
}

\begin{abstract}
In this paper, we study the periodic orbits and gravitational wave radiation in an extreme mass ratio inspiral system, where a stellar-mass object orbits a supermassive black hole without Cauchy horizons. Firstly, by using the effective potential, the marginally bound orbits and the innermost stable circular orbits are investigated. It is found that the radius, orbital angular momentum, and energy increase with the hair parameter for both orbits. Based on these results, we examine one special type of orbit, the periodic orbit, around the black hole without the Cauchy horizon. The results show that, for a fixed rational number, the energy and angular momentum of the periodic orbit increase with the hair parameter. In particular, we observe a significant deviation from the Schwarzschild case for small hair parameter with a large amount of external mass outside the black hole horizon. Moreover, we examine the waveforms in the extreme mass ratio inspiral system to explore the orbital information of the periodic orbits and the constraints on the parameters of the black holes. The results reveal that the gravitational waveforms can fully capture the zoom-whirl behavior of periodic orbits. Moreover, the phase of the gravitational waves imposes constraints on the parameters of the black hole solutions. As the system evolves, the phase shift of the waveforms becomes increasingly significant, with cumulative deviations becoming more pronounced over time. Compared to the Schwarzschild black hole background, the waveform phase will advance for the central supermassive black hole without a Cauchy horizon.
\end{abstract}



\maketitle

\section{Introduction}

Black hole and gravitational wave (GW), as two important theoretical predictions of general relativity, have been observationally confirmed to a significant extent through groundbreaking observations \cite{LIGOScientific:2016aoc, LIGOScientific:2016sjg, EventHorizonTelescope:2019dse, EventHorizonTelescope:2019uob}. Despite these achievements, general relativity continues to present unresolved challenges, such as the singularity problem, which remains a topic of deep investigation \cite{Penrose:1969pc, Hawking:1970zqf, Senovilla:1998oua}. Among the studies, GW radiation from the extreme mass ratio inspirals (EMRIs) system possesses rich information and can significantly enhance our understanding of the physical properties near the black hole event horizon, making them one of the most promising tools for probing the nature of black holes.

For EMRIs, the mass ratio between the two objects is highly unequal, with the smaller object rapidly orbiting the supermassive black hole. The orbit gradually decays over time, and GWs are emitted during this process. Due to the extremely small fraction of energy carried away by the lower-mass object's orbital motion compared to the system's total energy, the inspiral timescale of the smaller-mass object around the supermassive black hole can extend over several years. During this period, the object can complete approximately $10^3 \sim 10^5$  orbital cycles. The prolonged duration of the EMRI signals not only facilitates the accumulation of a higher signal-to-noise ratio \cite{Baiblack holeav:2021}, but also provides an extended observation window for precise parameter estimation in GW physics, thereby highlighting the significant scientific value of GWs emitted from EMRIs. From a fundamental physics perspective, since the GWs of EMRIs vary depending on the underlying gravitational theory, detecting these waveforms provides an approach to validate or challenge current models of gravity \cite{Amaro-Seoane:2019umn, Liang:2022gdk, AbhishekChowdhuri:2023gvu, Kumar:2024utz}. Thus, EMRIs offer an opportunity to test general relativity and probe modified theories of gravity. From an astronomical perspective, the detection of EMRIs can be used to accurately measure the intrinsic parameters of supermassive black holes, such as their mass, spin, and quadrupole moment \cite{Amaro-Seoane:2007osp}. Furthermore, the presence of dark matter surrounding a supermassive black hole could influence the motion of the compact object through dynamical friction, potentially altering the EMRI waveform. Therefore, by analyzing EMRI signals, we can also investigate the existence of dark matter around supermassive black holes \cite{Yue:2017iwc, Yue:2018vtk,  Duque:2023seg, Rahman:2023sof}. In the field of cosmology, EMRIs serve as standard sirens, providing precise constraints on cosmological parameters \cite{Schutz:1986gp}. Predictions for the occurrence rate of EMRI vary for different theoretical models, and most estimates suggest that future space-based GW detectors, such as LISA \cite{Danzmann:1997hm, Schutz:1999xj, Gair:2004iv, LISA:2017pwj, Maselli:2021men,Ghosh2024}, Taiji \cite{Hu:2017mde}, and Tianqin \cite{TianQin:2015yph}, will be capable of detecting several, or even a large number of EMRI events \cite{Babak:2017tow, Amaro-Seoane:2012lgq}.

On one hand, as one of the most anticipated research targets for upcoming space-based GW observatories, EMRIs provide a valuable wau to probe the nature of black holes and gravity, with broad implications for fundamental physics, astronomy, and cosmology. On the other hand, the characteristics of EMRI orbits, including eccentricity, orbital inclination, and highly relativistic motion, make the computation of the trajectories of compact objects and their corresponding GW forms particularly challenging. In EMRIs,  a compact object, such as a stellar-mass black hole or a neutron star, spirals inward toward a supermassive black hole, emitting GWs as a result of the inspiral. In this process, periodic orbits act as successive orbit transition states, playing a crucial role in understanding the GW radiation \cite{Glampedakis:2002ya}.
These periodic orbits provide key insights into the spacetime geometry near black holes, revealing valuable information about the behavior of matter and gravity in extreme conditions. Therefore, studying EMRIs through the periodic orbits and their gravitational radiations offers a powerful approach to explore the black hole spacetime in the strong-field regime.

Based on these facts, Levin \emph{et. al.} proposed a classification scheme for the periodic orbits of massive particles, which has played a significant role in understanding black hole merger dynamics  \cite{Glampedakis:2002ya, Levin:2008mq, Healy:2009zm, Grossman:2011im}. Periodic orbits are particularly important as they capture the fundamental characteristics of orbital dynamics around black holes, with typical black hole trajectories deviating only slightly from these orbits. Specifically, a periodic orbit is defined as one in which a test particle returns to its initial position after a finite time. By restricting the particle motion to the equatorial plane, the return of the particle to its initial position implies that the ratio of its motion in the $r$- and $\phi$-directions is a rational number $q$. Levin et al. proposed that this ratio could be expressed in terms of three integers $(z, w, v)$, where $q=\omega_{\phi}/\omega_r-1=w+v/z$ \cite{Levin:2008mq}. Here, $\omega_{r}$ and $\omega_{\phi}$ represent the particle motion frequencies in the $r$- and $\phi$-directions, respectively. The integers $z$, $w$, and $v$ correspond to the particle  \emph{zoom}, \emph{whirl}, and \emph{vertex} behaviors, respectively. Thus, each periodic orbit is associated with a rational number $q$, linking angular and orbital frequencies while characterizing the unique properties of the orbit \cite{Levin:2008mq}. This classification scheme has been applied to the study of periodic orbits in various black hole spacetimes, such as Schwarzschild black holes \cite{Levin:2008mq, Lim:2024mkb}, Kerr black holes \cite{Levin:2008ci, Levin:2008yp, Levin:2009sk, Grossman:2011ps}, charged black holes \cite{Misra:2010pu}, Kerr-Sen black holes \cite{Liu:2018vea}, and naked singularities \cite{Babar:2017gsg}. Additionally, studies of periodic orbits in other spacetime backgrounds are gradually expanding \cite{Azreg-Ainou:2020bfl, Wei:2019zdf, Deng:2020yfm, Zhou:2020zys, Deng:2020hxw, Lin:2021noq, Gao:2021arw, Zhang:2022zox, Wang:2022tfo, Zhang:2022psr, Lin:2023rmo, Yao:2023ziq, Tu:2023xab, Huang:2024oli, Yang:2024lmj, Jiang:2024cpe, Li:2024tld, Meng:2024cnq, Zhao:2024exh, Junior:2024tmi, Shabbir:2025kqh,Chan:2025ocy}.

Furthermore, during the inspiral stage, when two black holes within an EMRI system approach each other accompanied by GW emission,  periodic orbits serve as successive transitional states, playing a crucial role in understanding GW radiation \cite{Glampedakis:2002ya}. As described in Refs. \cite{Levin:2008mq, Grossman:2011im}, periodic orbits provide a method for rapidly calculating GW waveforms in EMRIs under the adiabatic approximation, which has been widely applied in recent studies \cite{Tu:2023xab, Yang:2024lmj} and extended to different gravitational backgrounds \cite{Huang:2024oli, Meng:2024cnq, Zhao:2024exh, Junior:2024tmi, Shabbir:2025kqh}, providing valuable material for testing gravitational theories.

In general relativity, black holes are intriguing entities defined by spacetime singularities hidden behind an event horizon \cite{Penrose:1969pc, Hawking:1970zqf, Senovilla:1998oua}. The existence of these singularities has motivated the exploration of alternative theories that go beyond general relativity, with the goal of preserving geodesic completeness in spacetime. In this context, a widely studied approach is to replace the singular region with a regular de Sitter space \cite{Sakharov:1966aja, Gliner:1966a, Frolov:1989pf}, a concept that, although introduced long ago, has recently experienced a resurgence of interest. Many studies in this field focus on constructing regular black hole solutions by modifying the static Misner-Sharp mass to ensure its rapid convergence to zero at small distances. This approach forms the basis for various static regular black hole models, including the Poisson-Israel model \cite{Poisson:1988wc}, the Dymnikova regular black hole solution \cite{Dymnikova:1992ux, Platania:2019kyx}, the asymptotically safe framework \cite{Bonanno:2000ep}, the Hayward metric \cite{Hayward:2005gi}, and regular black holes arising from pure gravity \cite{Bueno:2024dgm}, among others. For a comprehensive discussion, we refer to the detailed review in Ref. \cite{Bambi:2023}. Despite the regular nature of these black hole solutions, all of them feature a Cauchy horizon null hypersurface beyond which the predictability of spacetime dynamics breaks down. This leads to a phenomenon known as mass inflation at the perturbative level \cite{Poisson:1989zz, Poisson:1990eh}, which has been observed even in loop quantum gravity-inspired models \cite{Brown:2011tv}. Recent progress in understanding mass inflation has shown that singularity regularization does not always result in exponential instabilities \cite{Bonanno:2020fgp, Carballo-Rubio:2018pmi, Franzin:2022wai, Bonanno:2022jjp, Carballo-Rubio:2024dca, Khodadi:2024efq, Carballo-Rubio:2022kad, Bonanno:2023rzk, DiFilippo:2022qkl}.
Nevertheless, the presence of the Cauchy horizon continues to pose significant challenges \cite{Carballo-Rubio:2018pmi}. One potential solution is the construction of regular black holes that entirely lack an inner horizon \cite{Carballo-Rubio:2024dca, Casadio:2022ndh}.

This paper focuses on investigating the periodic orbits around black holes without Cauchy horizons \cite{Ovalle:2023vvu}, which are constructed within the framework of general relativity by applying the gravitational decoupling method to linearize the mass function in the metric. Specifically, we consider black holes without Cauchy horizons and with integrable singularities as the central supermassive black hole, while treating stellar-mass objects as test particles to form extreme mass ratio inspirals (EMRIs). By analyzing the trajectories of periodic orbits, we aim to study the gravitational waveforms emitted by EMRIs, providing insights into the gravitational radiation generated in such systems. In addition to the analysis of periodic orbits, studies have also been carried out for the black holes without Cauchy horizons in cases involving charge \cite{Tello-Ortiz:2024oih}, $d$-dimensional spacetime \cite{Estrada:2023dcj, Estrada:2024moz}, and rotation \cite{Leon:2024uao}.

The structure of the paper is as follows. In Sec. \ref{black holeandGEO}, we first introduce the black holes without Cauchy horizons and then study their geodesics. In Sec. \ref{periodicorbits}, we focus on the study of periodic orbits characterized by the rational number $q$, expressed in terms of three integers around the black holes without Cauchy horizons. This section also incorporates the classification of bound orbits based on the zoom-whirl structure \cite{Levin:2008mq}. In Sec. \ref{GWandpo}, we turn to the GW radiation emitted by periodic orbits around black holes without Cauchy horizons. Finally, in Sec. \ref{Conclusion}, we summarize and discuss our results.

\section{BLACK HOLE without Cauchy horizon AND GEODESICS}
\label{black holeandGEO}

In this section, we first provide a brief introduction to black holes without Cauchy horizons, then examine the geodesic motion of massive particles around the black hole.

\subsection{The construction of Black hole without Cauchy horizon}

In order to consider the black holes without Cauchy horizons, one can start with the following spherically symmetric static spacetime \cite{Ovalle:2023vvu, Visser:1995}
\begin{eqnarray}\label{lineelement}
ds^2&=&-e^{\Phi(r)}\bigg[1-\frac{2m(r)}{r}\bigg]dt^2+\frac{dr^2}{1-\frac{2m(r)}{r}}\nonumber\\
&&+r^2(d\theta^2+\sin^2\theta d\phi^2).
\end{eqnarray}
The function $\Phi(r)$ represents the metric function, while $m(r)$ is the Misner-Sharp mass function, which quantifies the total energy contained within a spherical region of radius $r$ and determines the mass distribution of the black hole system. When the metric function $\Phi(r)=0$, the line element \eqref{lineelement} reduces to the Kerr-Schild class metric \cite{Kerr:1965wfc}, which has been extensively studied (see, for example, Ref. \cite{Jacobson:2007tj}). One characteristic of the Kerr-Schild metric is that the mass function in the field equation takes a linear form with respect to the derivative of $r$, which allows multiple mass distributions to be superimposed \cite{Ovalle:2023vvu}. The core idea of Ref. \cite{Ovalle:2023vvu} is to eliminate the Cauchy horizon and ensure the integrability of singularities by constructing an appropriate mass function $m(r)$: (a) This ``gravitational decoupling'' technique \cite{Ovalle:2017fgl, Ovalle:2018gic}, which ensures the mass function $m^{\prime}(r)$ satisfies linearity conditions, allows the black hole solution to simultaneously satisfy both the non-singularity condition and the energy condition. (b) The original solution $m(r)$ is modified to the new solution by adding a correction term $\hat{m}(r)$,
\begin{equation}\label{massfun}
 m(r)\to\bar{m}(r)=m(r)+\hat{m}(r).
\end{equation}

Here, we briefly review the construction of the mass function in Ref.  \cite{Ovalle:2023vvu}. A key requirement is the continuity of the metric \eqref{lineelement} with $\Phi(r)=0$. In the radial direction, the stellar object is divided into two parts by $r_s$. If this describes the spacetime of a stellar object with radius $r = r_s$ in both its inner and outer regions, then, for a smooth matching of the two regions at $r = r_s$, the mass function $m(r)$ must satisfy:
\begin{equation}\label{continue}
m(r_s) = \tilde{m}(r_s)  \qquad {\rm and}\qquad m^{\prime}(r_s) = \tilde{m}^{\prime}(r_s),
\end{equation}
concurrently, where $\tilde{m}$ denotes the exterior mass function, and $F(r_s) \equiv F(r)|_{r=r_s}$ for any function $F(r)$.

In the case of the metric \eqref{lineelement} with $\Phi(r)=0$, in order to eliminate the Cauchy horizon and satisfy the mass function shown in Eq.~\eqref{massfun}, the mass function provided in Ref. \cite{Ovalle:2023vvu} for the region $r<r_s$ is given by
\begin{equation}\label{innermassfunr}
	m(r)=\frac {r}{2}\left[1+\left[1-\left(\frac r{r_s}\right)^n\right]^k\right];\qquad  r\leq r_s,
\end{equation}
where $n$ and $k$  are free parameters that control the properties of the mass function.

It should be noted that the metric generated by the mass function \eqref{innermassfunr} is not globally defined but is confined to the region $r\leq r_s$. The total mass $M$ of the system within $r<r_s$ is defined as $M\equiv m(r_s)=r_s/2$, independent of the parameters $k$ and $n$. Therefore, the event horizon is equivalent to the Schwarzschild radius, with $r_s=2M$. In addition, different values of the parameters $k$ and $n$ can give rise to a wide range of interesting and diverse physical environments. For instance, when $k=1$, the metric derived from the mass function \eqref{innermassfunr} satisfies the strong energy condition, while for $k>6$, it satisfies the dominant energy condition \cite{Ovalle:2023vvu}. In Ref. \cite{Ovalle:2023vvu}, considering the presence of a de Sitter core in regular black holes, only the case with $n=2$ is examined. Specifically, for ${k=1,n=2}$, it ensures the integrability of the scalar
curvature $ R=-4\Lambda+4/r^2$, meaning that the curvature remains finite. In the case of $n=2$, the parameter $k$ determines the degree of deformation of spacetime inside the object $(r<r_s)$. When $k>1$, the deviation of the internal spacetime from anti-de Sitter space can be measured. Specifically, when $k$ takes integer values, the internal spacetime can be interpreted as a finite superposition of multiple anti-de Sitter spaces. This also serves as an important manifestation
of the linearity of the Kerr-Schild mass function, which is given in Eq. (16) of Ref. \cite{Ovalle:2023vvu}.

Kerr-Schild spacetime requires the continuity of both the Einstein field equations and the mass function across the boundary $r_s$, ensuring that the energy density and radial pressure remain continuous both inside and outside $r_s$. However, as shown in Ref. \cite{Ovalle:2023vvu}, the energy density $\epsilon$  and radial pressure $p_r$ in the internal solution at $r_s$ are not zero. Consequently, the external solution at  $r_s$ cannot be smoothly matched with the internal solution using the Schwarzschild (vacuum) solution. Therefore, the author of Ref. \cite{Ovalle:2023vvu} proposed a new external solution to facilitate the transition between the internal solution and the Schwarzschild solution, allowing for a rapid approach to the external Schwarzschild solution. To achieve this, the author employs the so-called gravitational decoupling technique \cite{Ovalle:2020kpd}, which satisfies the strong energy condition for the metric. The corresponding mass function is given by:
\begin{equation}\label{outermassfun1}
\tilde{m}(r)=\mathcal{M}-\alpha \frac{r}{2}e^{-2r/(2 \mathcal{M}-\ell)};\qquad r\geq r_s,
\end{equation}
where $\mathcal{M}$ represents the asymptotic mass, the total mass measured by an observer at infinity, $\alpha$ quantifies the deviation from the Schwarzschild solution, and $\ell$ is a ``hair'' parameter, which currently satisfies $0 < \ell \leq 2\mathcal{M}$ to ensure asymptotic flatness. It is important to emphasize that $\mathcal{M}$ and $M$ are distinct: $\mathcal{M}$ refers to the total mass of the entire configuration, while $M$ denotes the portion of the total mass confined within the region $r\leq r_s$.

By applying the matching condition \eqref{continue} to connect the mass functions \eqref{innermassfunr} and \eqref{outermassfun1} at $r=r_s$, the external mass function is ultimately obtained as
\begin{equation}\label{outermassfun2}
\tilde{m}(r)=\mathcal{M}+\frac{r}{2}\left( 1-\frac{2\mathcal{M}}{r_s}\right)e^{-2(r-r_s)/(2 \mathcal{M}-\ell)};\qquad r\geq r_s.
\end{equation}
Substituting the mass function \eqref{outermassfun2} into the metric \eqref{lineelement} with $\Phi(r) = 0$, we obtain the metric for the region $r\geq r_s$ as follows:
\begin{equation}\label{metric}
ds^2=-f(r)dt^2+\frac{1}{f(r)}dr^2+r^2 d\Omega^2;\qquad r\geq r_s,
\end{equation}
with the lapse function $f(r)$ being given by \cite{Ovalle:2023vvu}
\begin{equation}\label{lapsefunction}
f(r)=1-\frac{2 \mathcal{M}}{r}-\left(1-\frac{2 \mathcal{M}}{r_{\mathrm{s}}}\right) e^{-2\left(r-r_{\mathrm{s}}\right) /(2 \mathcal{M}-\ell)}.
\end{equation}
From the metric \eqref{metric}, it is clear that $r_s$  represents the event horizon of the black hole in the metric constructed by the external mass function
\begin{equation}\label{horizon}
r_h=r_{\mathrm{s}}=2 M=\left(\mathcal{M}+\sqrt{\ell \mathcal{M}-\mathcal{M}^{2}}\right).
\end{equation}
Thus, $r_s$ serves two purposes: on the one hand, it represents the location where the metric is smoothly connected through the inner and outer mass functions; on the other hand, $r_s$ corresponds to the event horizon $r_h$ of the metric constructed from the external mass function in Eq. \eqref{lapsefunction}. Moreover, the Cauchy horizon $r_{h_{(i)}}$, which forms the other branch of the metric constructed from the external mass function in Eq. \eqref{lapsefunction}, is removed by the metric constructed from the internal mass function. From Eq.~\eqref{horizon}, it can be seen that $\mathcal{M}$ tends to $M$ as $\ell \to \mathcal{M}$, meaning that $\ell$ controls the amount of mass $\mathcal{M}$ contained within the smooth connection surface $r\leq r_s$ of the inner and outer mass functions. However, according to the continuity of energy density and radial pressure across the Kerr-Schild spacetime (e.g., see Eq. (9) in Ref. \cite{Ovalle:2023vvu}), the solution corresponding to the external mass function cannot be a Schwarzschild solution. Therefore, we have $\mathcal{M}\leq \ell < 2\mathcal{M}$. Interestingly, when $\ell=\mathcal{M}$, the masses inside and outside $r_s$ are equal (namely, $M=\mathcal{M}$), representing an extreme case. Regardless, when $\ell$ is just slightly less than $2\mathcal{M}$, meaning that the black hole's event horizon still covers some mass outside, the solution remains valid. In Ref. \cite{Ovalle:2023vvu}, it is pointed out that the external mass outside the black hole is exactly zero, representing a vacuum situation (Schwarzschild case), which is an idealized case. Of course, from a phenomenological perspective, the external mass could be attributed to components such as an accretion disk. It also mentions that the larger the external mass ($\ell$ smaller), the more tolerant the solution is to the singularities.

\subsection{The properties of black hole without Cauchy horizon}

\begin{figure*}[htbp]
\centering
	\subfigure[$f(r)-r/\mathcal{M}$.]{\label{frr}
	\includegraphics[width=6.5cm]{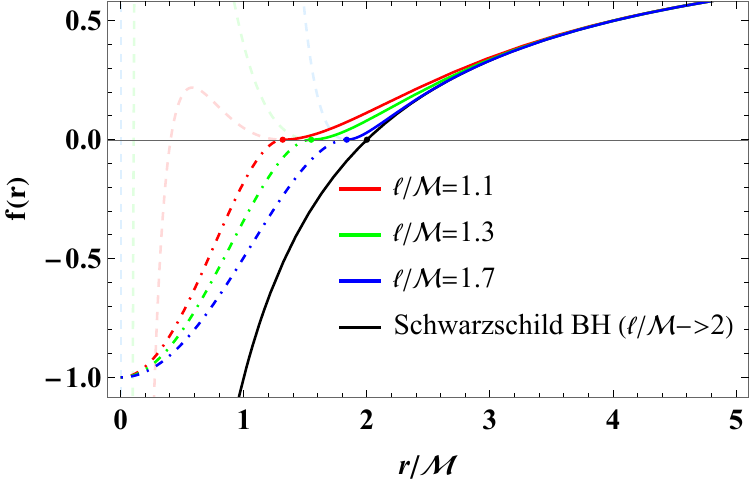}}
	\subfigure[~$r/\mathcal{M}-\ell/\mathcal{M}$.]{\label{rlPara}
	\includegraphics[width=6.5cm]{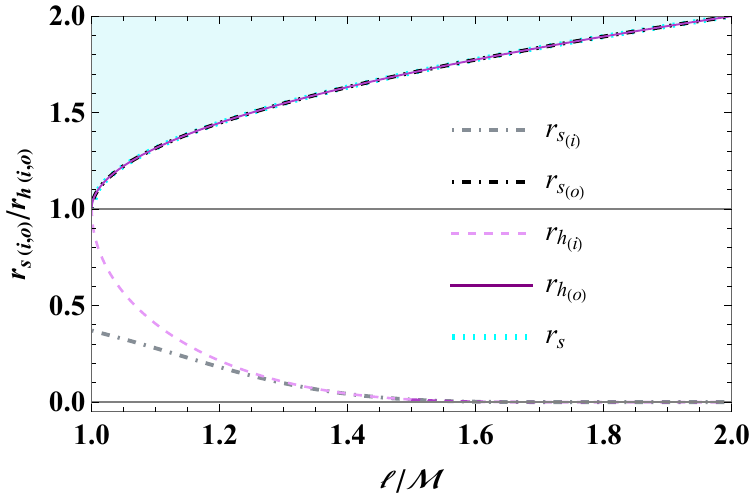}}
\captionsetup{justification=raggedright}
\caption{(a) The behavior of the metric function $f(r)$ constructed from the internal $m(r)$ and external $\tilde{m}(r)$ mass functions.  (b) The dependence of the smooth connection radius $r_{s_{(i,o)}}$ and the horizons $r_{h_{(i,o)}}$ on the dimensionless hair parameter $\ell/\mathcal{M}$.}
\label{rlParameter}
\end{figure*}

To gain a clearer understanding of black holes without Cauchy horizons, we use Fig.~\ref{rlParameter} to further illustrate some features of the black hole solution. In Fig.~\ref{frr}, the curves of the metric function $f(r)$ generated by the inner $m(r)$ and outer $\tilde{m}(r)$ mass functions are shown. The dot-dashed line represents $f(r)$ generated by $m(r)$, which replaces the dashed line in light color representing $f(r)$ generated by $\tilde{m}(r)$ inside $r_s$, thus achieving a configuration without a Cauchy horizon. The internal mass function $m(r)$ is used to construct the metric when $k=2$ and $n=2$, where the choice of $k$ does not affect the external mass function. The solid line shows the behavior of $f(r)$ generated by $\tilde{m}(r)$ outside the event horizon. The points on the curve represent the location of the event horizon.

In Fig.~\ref{rlPara}, the behaviors of the connection point $r_s$ and the black hole horizon with varying hair parameter $\ell$ are shown. The black dot-dashed lines represent the two branches of solutions to Eq.~\eqref{continue}, where the inner and outer mass functions correspond to $m(r)$ in Eq.~\eqref{innermassfunr} and $\tilde{m}(r)$ in Eq.~\eqref{outermassfun2}, respectively. The gray dot-dashed line denotes the inner solution $r_{s_{(i)}}$, while the dark black dot-dashed line represents the outer solution $r_{s_{(o)}}$, which is also the horizon derived from $f(r) = 0$ using the inner mass function $m(r)$. The purple lines correspond to the two branches of Eq.~\eqref{lapsefunction}: the external dark purple solid line represents the event horizon $r_{h_{(o)}}$ (same as Eq.~\eqref{horizon}), and the internal light purple dashed line represents the discarded ``Cauchy horizon'' $r_{h_{(i)}}$. The cyan dotted line marks the event horizon constructed by the inner and outer mass functions, coinciding with $r_{s_{(o)}}$. Outside this region lies the true exterior of the black hole. Here, the total mass $M$ is contained within $r_s$, but for $1\leq \ell/\mathcal{M}<2$, part of $\mathcal{M}$ is distributed on the surface $r_s$.

From Fig.~\ref{rlPara}, it can be seen that when the inner and outer mass functions are smoothly joined, the horizons $r_h$ at the connecting boundary $r_{s_{(i,o)}}$ is highly sensitive to the hair parameter $\ell/\mathcal{M}$. As the hair parameter $\ell/\mathcal{M}$ increases, the outer horizon $r_{h_{(o)}}$ and the outer solution $r_{s_{(o)}}$ move outward, approaching the Schwarzschild solution's $r_g=2\mathcal{M}$. The ``inner horizon'' $r_{h_{(i)}}$ and $r_{s_{(i)}}$ move inward toward the coordinate origin. As $\ell$  approaches $2\mathcal{M}$, the behavior of Eq.~\eqref{horizon} converges to the event horizon of the Schwarzschild black hole background. Certainly, from Fig.~\ref{rlParameter}, we can observe that when the external mass is very small (with a relatively large hair parameter), both the inner branches of the solutions from Eqs. \eqref{continue} and \eqref{horizon} tend toward zero. This indicates that the function properties remain well-behaved in this case, which may reflect the true state of the black hole. In Ref. \cite{Ovalle:2023vvu}, the black holes without Cauchy horizons are constructed. This challenges the conventional view on the existence of singularities and Cauchy horizons in traditional black hole models, offering a new perspective for understanding the structure and properties of black holes. By investigating $\ell$ through periodic orbits and GW radiation, we can infer the mass function distribution inside and outside $r_s$ for black holes without Cauchy horizons, allowing for the study of properties such as energy density distribution. This could provide important information for understanding the interactions between black holes and their environment, radiation properties, and the long-term evolution of black holes.

\subsection{Geodesics and effective potential of the black hole with Cauchy horizon}

In this subsection, we consider the motion of a test particle around the black hole without the Cauchy horizon. For a free particle, its Lagrangian is given by
\begin{equation}\label{Lagrangian}
\mathcal{L}=\frac{1}{2}g_{\mu\nu}\dot{x}^{\mu}\dot{x}^{\nu}=\delta,
\end{equation}
with $\delta=- 1$ and 0 for massive and massless particles, respectively. The dot here represents the derivative with respect to the particle's proper time $\tau$. Due to the spherical symmetry of spacetime, we consider the equatorial motion with $\theta=\pi/2$ and $\dot{\theta}=0$ without loss of generality. According to the Lagrangian, the generalized momentum of the particle is:
\begin{equation}\label{genmom}
p_{\mu}=\frac{\partial \mathcal{L}}{\partial \dot{x^{\mu}}}.
\end{equation}
Then the components of the momentum are
\begin{gather}\label{momentum}
p_{t}=g_{tt}\dot{t}=-E,\\
p_{\phi}=g_{\phi\phi}\dot{\phi}=L,\\
p_{r}=g_{rr}\dot{r},\\
p_{\theta}=g_{\theta\theta}\dot{\theta}=0.
\end{gather}
For each geodesic, there are two corresponding constants, $E$ and  $L$, which represent the conservation of energy and the orbital angular momentum per unit mass, respectively. Combining the above three equations with the normalization condition for the four-velocity of a massive particle, $g_{\mu\nu}\dot{x^{\mu}}\dot{x^{\nu}}=-1$, we obtain
\begin{gather}
\dot{t}=-\frac{E}{f(r)}, \label{tdot}\\
\dot{\phi}=\frac{L}{r^2}, \label{phidot}\\
\dot{r}^2=E^2-f(r)\left(1+\frac{L^2}{r^2}\right) \label{rdot1}.
\end{gather}
Obviously, the geodesic motion around the black hole without the Cauchy horizon depends on the hair parameter $\ell$ in $f(r)$, which characterizes the deviation from the Schwarzschild background.

Since the particle is moving on the equatorial plane, we only need to consider motion in the radial $r$ direction and the angular $\phi$ direction. First, for simplicity, we rewrite Eq.~\eqref{rdot1} as
\begin{equation}\label{rdot2}
\dot{r}^2=E^2-V_{\mathrm{eff}},
\end{equation}
where $V_{\mathrm{eff}}$ denotes the effective potential
\begin{equation}\label{effpot}
V_{\mathrm{eff}}=f(r)\left(1+\frac{L^2}{r^2}\right).
\end{equation}
It is easy to verify that, under the condition $\mathcal{M}>0$ and $\mathcal{M}\leq \ell<2\mathcal{M}$, $V_{\mathrm{eff}}\to 1$ as $r\to +\infty$. Taking appropriate conditions, we shall obtain the marginally bound orbit (MBO) and the innermost stable circular orbit (ISCO).

\subsection{Marginally bound orbits}

\begin{figure*}[htbp]
	\centering
	\subfigure[~$r_{\rm MBO}$ vs $\ell$.]{\label{rMBO2a}
	\includegraphics[width=6.5cm]{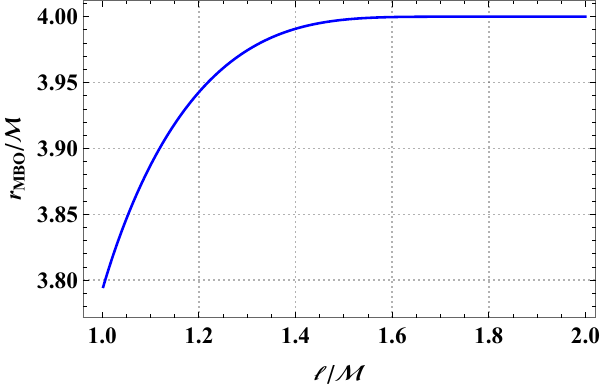}}
	\subfigure[~$L_{\rm MBO}$ vs $\ell$.]{\label{LMBO2b}
	\includegraphics[width=6.5cm]{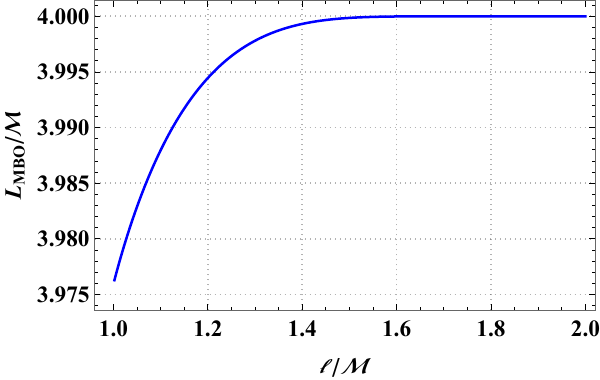}}
\captionsetup{justification=raggedright}
\caption{The radius $r_{\rm MBO}$ and angular momentum $L_{\rm MBO}$ of MBOs as a function of the hair paraemter $\ell/\mathcal{M}$. }
\label{fig:rLMBO}
\end{figure*}

Let us first consider a marginally bound orbit, which is defined as follows:
\begin{equation}\label{mbo}
V_{\mathrm{eff}}=E^2,\qquad\partial_{r}V_{\mathrm{eff}}=0.
\end{equation}
By inserting Eq.~\eqref{effpot} into (\ref{mbo}), we can numerically solve for the radius $r_{\rm MBO}$  and orbital angular momentum $L_{\rm MBO}$ of the MBO for a fixed hair parameter $\ell$. The results are shown in Fig.~\ref{fig:rLMBO}. Since the metric function $f(r)$ is based on the Schwarzschild solution modified by an exponential function, it can be observed that as the hair parameter increases, the $r_{\rm MBO}$ and $L_{\rm MBO}$ of the MBO rapidly approach the Schwarzschild case.

\subsection{Innermost stable circular orbits}

Now, let us turn to consider the ISCO, which is defined as
\begin{equation}\label{Isco}
V_{\mathrm{eff}}=E^2,\qquad\partial_{r}V_{\mathrm{eff}}=0,\qquad\partial_{r,r}V_{\mathrm{eff}}=0.
\end{equation}
Since the metric is spherically symmetric, we can obtain the radius $r_{\mathrm{ISCO}}$ of ISCO by solving \eqref{Isco}, and the expressions for the angular momentum $L_{\mathrm{ISCO}}$ and energy $E_{\mathrm{ISCO}}$ of the orbit are given as follows
\begin{gather}\label{rLEIsco}
r_{\mathrm{ISCO}}=\frac{3 f\left(r_{\mathrm{ISCO}}\right) f^{\prime}\left(r_{\mathrm{ISCO}}\right)}{2 f^{\prime 2}\left(r_{\text {ISCO}}\right)-f\left(r_{\text {ISCO}}\right) f^{\prime \prime}\left(r_{\text {ISCO}}\right)}, \\
L_{\mathrm{ISCO}}=r_{\mathrm{ISCO}}^{3 / 2} \sqrt{\frac{f^{\prime}\left(r_{\mathrm{ISCO}}\right)}{2 f\left(r_{\mathrm{ISCO}}\right)-r_{\mathrm{ISCO}} f^{\prime}\left(r_{\mathrm{ISCO}}\right)}},\\
E_{\mathrm{ISCO}}=\frac{f\left(r_{ISCO}\right)}{\sqrt{f\left(r_{\mathrm{ISCO}}\right)-r_{\mathrm{ISCO}} f^{\prime}\left(r_{\mathrm{ISCO}}\right)}}.
\end{gather}
\begin{figure*}[htbp]
	\center{
	\subfigure[~$r_{\rm ISCO}$ vs $\ell$. ]{\label{risco3a}
	\includegraphics[width=5.5cm]{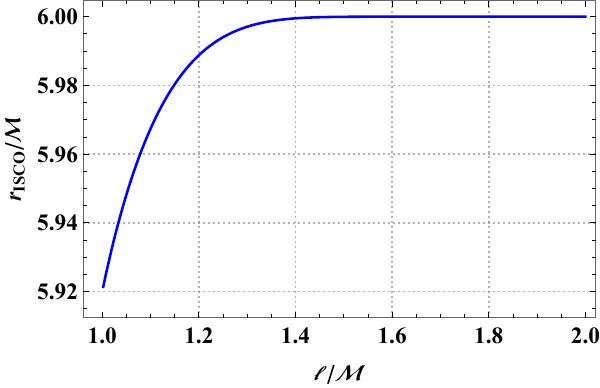}}
	\subfigure[~$L_{\rm ISCO}$ vs $\ell$. ]{\label{Lisco3b}
	\includegraphics[width=5.5cm]{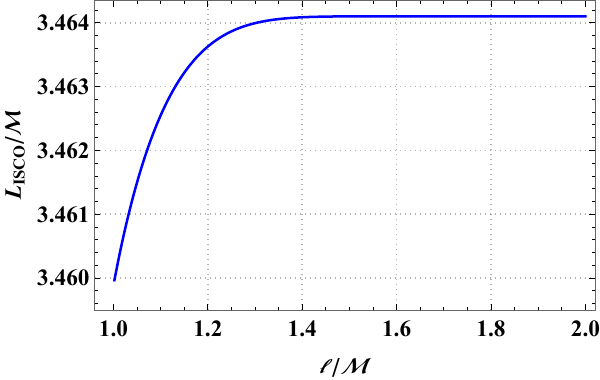}}
	\subfigure[~$E_{\rm ISCO}$ vs $\ell$.]{\label{Esco3c}
	\includegraphics[width=5.5cm]{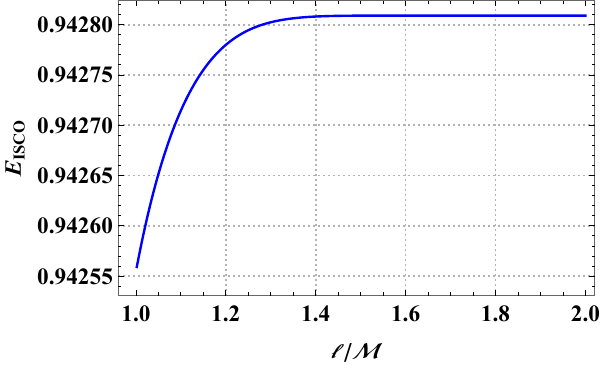}}
	}
\captionsetup{justification=raggedright}
\caption{The radius $r_{\rm ISCO}$, energy $E_{\rm ISCO}$, and angular momentum $L_{\rm ISCO}$ of ISCOs.}
\label{fig:rLEISCO}
\end{figure*}
In Fig.~\ref{fig:rLEISCO}, we plot the result of $r_{\mathrm{ISCO}}$, $L_{\mathrm{ISCO}}$ and $E_{\mathrm{ISCO}}$ with respect to the hair parameter $\ell$ of the black hole without the Cauchy horizon. Their behaviors are similar to that shown in Fig.~\ref{fig:rLMBO}, both of which rapidly approach the Schwarzschild case as the hair parameter $\ell$ increases.

\begin{figure*}[htbp]
	\center{
	\subfigure[$\ell/{\mathcal{M}}=1 $]{\label{veff4a}
	\includegraphics[width=6.5cm]{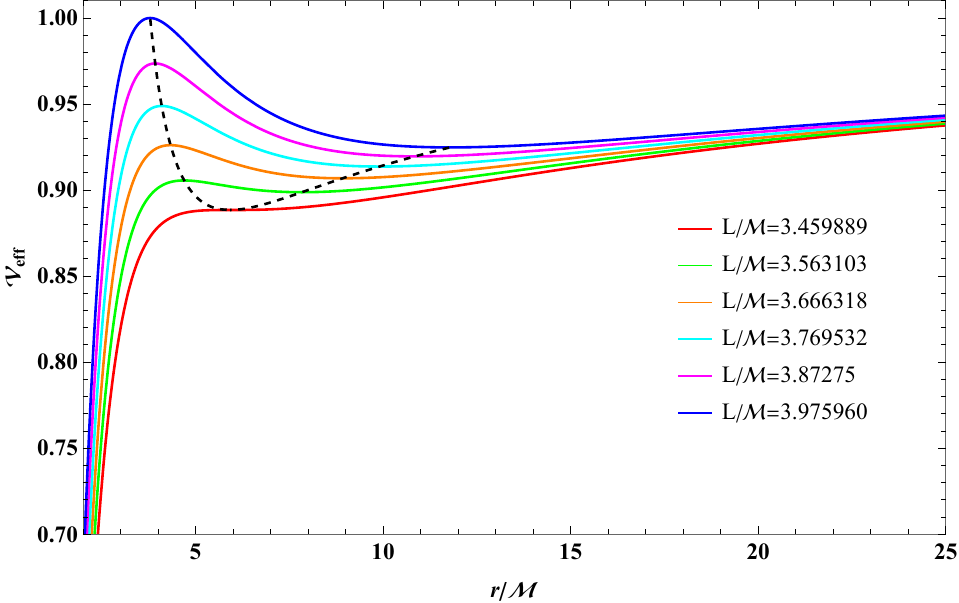}}
	\subfigure[$\ell/{\mathcal{M}}=1.99 $]{\label{veff4b}
	\includegraphics[width=6.5cm]{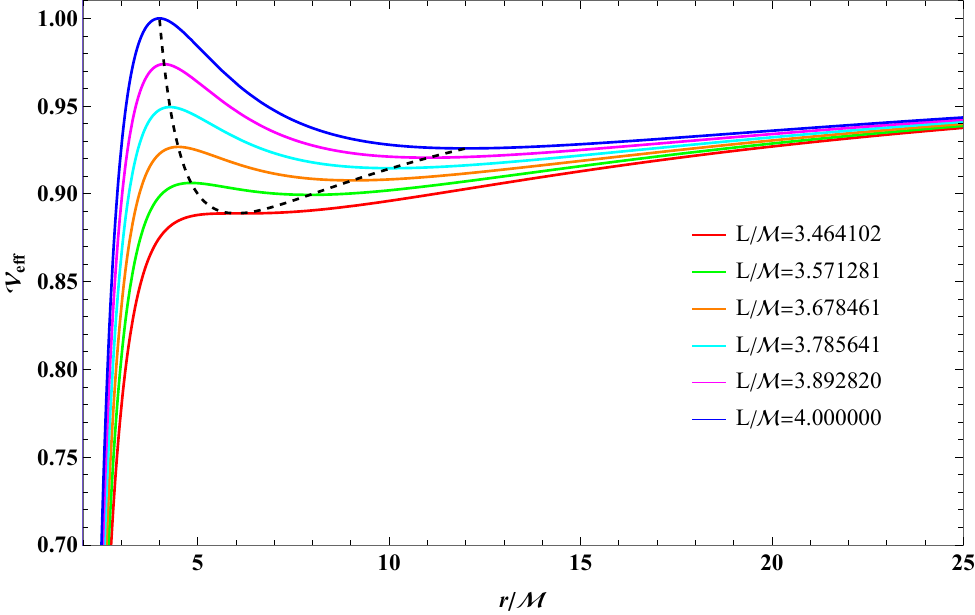}}
	}
\captionsetup{justification=raggedright}
\caption{The effective potential $V_{\mathrm{eff}}$ as a function of $r/\mathcal{M}$. The angular momentum $L/\mathcal{M}$ varies from $L_{\rm ISCO}$ to $L_{\rm MBO}$ from bottom to top. The dashed line represents the extremal points of the effective potential. The black hole horizon is located at $r_s=1 \mathcal{M}$ and $r_s=1.995  \mathcal{M}$, respectively.}
\label{fig:potential}
\end{figure*}

In Fig.~\ref{fig:potential}, we present the graphs of the effective potential as a function of the radial coordinate $r/\mathcal{M}$ for $\ell/\mathcal{M}=1$ and $\ell/\mathcal{M}=1.99$, respectively. The curves from bottom to top show that the orbital angular momentum changes from $L_{\rm ISCO}$ to $L_{\rm MBO}$. The extremum points of the effective potential are marked by black dashed lines. It can be observed that as the radial radius decreases, the extremum of the effective potential shifts. When $\partial_{r,r} V_{\mathrm{eff}}=0$, the effective potential has only one extremum, at which the radius corresponds to the ISCO, and we denote the orbital angular momentum as $L_{\rm ISCO}$. When the orbital angular momentum increases, the effective potential exhibits two extrema, corresponding to the inner unstable orbit and the outer stable orbit. In Fig.~\ref{veff4a}, it can be seen that when deviating from the Schwarzschild solution, the corresponding values are $r_{\rm ISCO}=5.9201\mathcal{M},~L_{\rm ISCO}=3.4599\mathcal{M}$, and $E_{\rm ISCO}=0.9425$. In contrast, in Fig.~\ref{veff4b}, when approaching the Schwarzschild solution, the values are $r_{\rm ISCO}=5.9999\mathcal{M},~L_{\rm ISCO}=3.4641\mathcal{M}$, and $E_{\rm ISCO}=0.9428$. The slight differences between the two sets of values are helpful in explaining the variations in the GW radiation and periodic orbits corresponding to the black hole without the Cauchy horizon and Schwarzschild black hole.

\begin{figure*}[htbp]
	\center{
	\includegraphics[width=6.5cm]{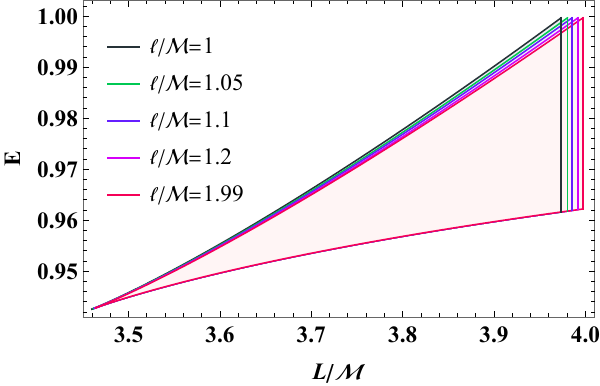}
	}
\captionsetup{justification=raggedright}
\caption{Parameter regions for the bound orbits (in shadow) for different $\ell/\mathcal{M}$.}
	\label{fig:ELpara}
	\end{figure*}

In addition, the parameter space of orbital angular momentum and energy corresponding to bound orbits can be studied by $\dot{r}=0$ and $\partial_r \dot{r}=0$. In Fig.~\ref{fig:ELpara}, we present the $E-L/\mathcal{M}$  parameter space for different  $\ell/\mathcal{M}$. It can be observed that when $\ell/\mathcal{M}$ is small, the range of allowable orbital angular momentum is relatively narrow, while the corresponding energy range is quite wide. Conversely, when $\ell/\mathcal{M}$ is large, the situation is reversed, with a wider range of allowed angular momentum and a narrower range of energy.

\section{PERIODIC ORBITS}
\label{periodicorbits}

A periodic orbit is a special type of bound orbit, which returns precisely to its initial position after a finite amount of time. Since the background is that of a spherically symmetric black hole, with $\theta=\pi/2$ and $\dot{\theta}=0$, the periodic orbit is entirely determined by the oscillation frequencies of the radial $r$-motion and the angular $\phi$-motion. Specifically, periodic orbits require that the ratio of these two frequencies be a rational number. According to Ref. \cite{Levin:2008mq}, one can define a rational number $q$
\begin{equation}\label{rationalnumber}
q=\frac{\omega_{\phi}}{\omega_{r}}-1=\frac{\Delta\phi}{2\pi}-1.
\end{equation}
For a periodic orbit, the radial period can be regarded as twice the time that the particle used moving between the two turning points, periapsis, and apoapsis, for that the particle must return precisely to its initial position within one period. $\Delta \phi$ refers to the angle by which the particle is deflected in the $\phi$-direction on the equatorial plane after one full radial cycle. It must also be an integer multiple of $2\pi$. The deviation angle $\Delta \phi$ in the $\phi$-direction over one radial period can be expressed as follows:
\begin{equation}\label{deltaphi}
{{\Delta\phi=\oint\mathrm{d}\phi}}\\ {{{}=2\int_{r_{p}}^{r_{a}}\frac{\dot{\phi}}{\dot{r}}\mathrm{d}r}}\\ {{{}=2\int_{r_{p}}^{r_{a}}{\frac{L}{r^2\sqrt{E^2-f(r)(1+\frac{L^2}{r^2})}}}}}.
\end{equation}
The factor of 2 arises because the particle's complete radial period involves traveling from apoapsis to periapsis and back, making the total distance twice the separation between the two turning points. By substituting \eqref{deltaphi} into \eqref{rationalnumber}, it can be seen that the factors influencing the rational number of periodic orbits are the particle's energy $E$, its orbital angular momentum $L$, and the metric function $f(r)$ of the black hole systems. For a bound orbit, the angular momentum $L$ only varies between $L_{\rm ISCO}$ and $L_{\rm MBO}$. To simplify the analysis and calculation, the angular momentum $L$ for a given bound orbit can be expressed in the following form:
\begin{equation}\label{angmom}
L=L_{\mathrm{ISCO}}+\epsilon(L_{\mathrm{MBO}}-L_{\mathrm{ISCO}}).
\end{equation}
The parameter $\epsilon$ is constrained to the range $(0,1)$, where $\epsilon=0$ and 1 correspond to the orbital angular momenta at the ISCO and MBO, respectively. When $\epsilon > 1$, bound orbits no longer exist. Therefore, by assigning different values to $\epsilon$, the orbital angular momentum can be specified.

On the other hand, in the context of periodic orbits, the rational number $q$ can be expressed as the ratio of three integers \cite{Levin:2008mq}:
\begin{equation}
q=w+\frac{v}{z}.
\end{equation}	
The integers $(z, w, v)$ have specific geometric interpretations. They are referred to as the zoom number $z$, the whirls number $w$, and the vertex number $v$, respectively. In the zoom-whirl classification, these integers represent the scaling, rotation, and vertex behaviors of the orbit. The number $q$ measures the extent of periapsis precession beyond a simple elliptical orbit, providing insight into the orbit's topology. Additionally, the tracing order of leaves, referring to the sequence of
orbital paths or segments, is also considered in this classification system. This approach helps in characterizing the complex dynamics of the periodic orbits.

\begin{figure*}[htbp]
	\center{
	\subfigure[$\ell/{\mathcal{M}}=1 $]{\label{Eq6a}
	\includegraphics[width=8cm]{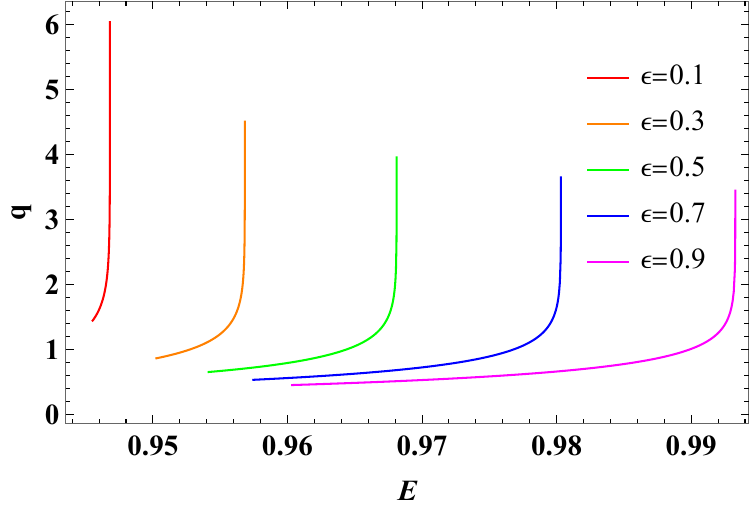}}
	\subfigure[$\epsilon=0.5 $]{\label{Eq6b}
	\includegraphics[width=8cm]{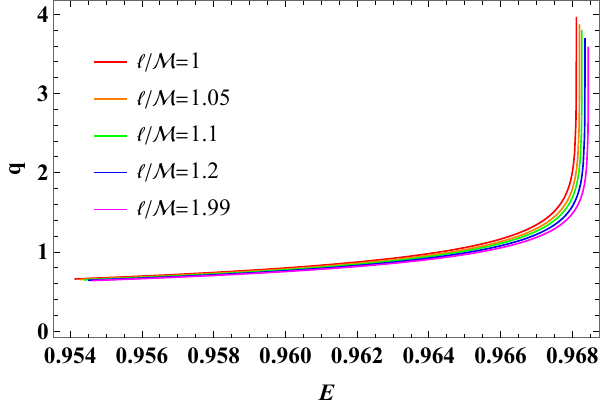}}\\
	\subfigure[$\ell/{\mathcal{M}}=1 $]{\label{Lq6c}
	\includegraphics[width=8cm]{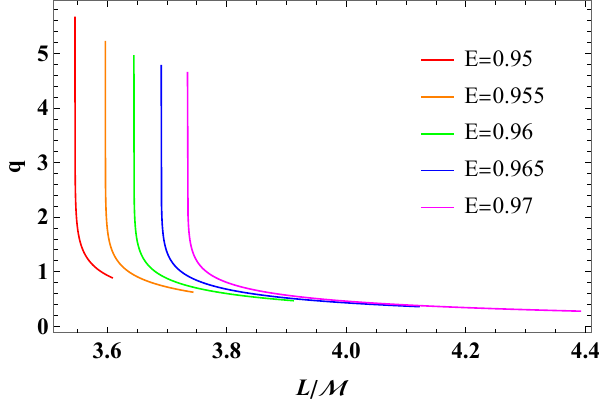}}
	\subfigure[$E=0.96 $]{\label{Lq6d}
	\includegraphics[width=8cm]{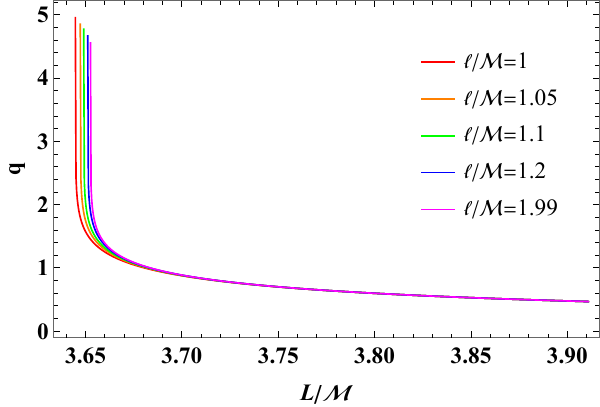}}
	}
\captionsetup{justification=raggedright}
\caption{The rational number $q$ as a function of the energy  (top panel) and angular momentum  (bottom panel) of periodic orbits around black holes without Cauchy horizons.}
\label{fig:qbehaviour}
\end{figure*}

To illustrate the behavior of $q$, we present $q$ versus particle energy $E$ and angular momentum $L/\mathcal{M}$ in Fig.~\ref{fig:qbehaviour}. In Figs. \ref{Eq6a} and \ref{Eq6b}, the variations of the rational number $q$ with particle energy are shown. It is found that $q$ increases slowly with energy $E$ initially and then increases rapidly after reaching the maximum energy. In Fig.~\ref{Eq6a}, the variation of $q$ with $\epsilon$ is shown for the case where the hair parameter $\ell/\mathcal{M}=1$ and $\epsilon$ takes different values. It is observed that $\epsilon$ has a significant impact on the distribution of \(q-E\). In Fig.~\ref{Eq6b}, we fix \(\epsilon = 0.5\) and explore the influence of different $\ell/\mathcal{M}$ on $q$. The results show that the closer the system is to the Schwarzschild case, the larger the maximum energy corresponding to $q$. On the other hand, in Figs. \ref{Lq6c} and \ref{Lq6d}, the behaviors of $q$ as a function of the  $L/\mathcal{M}$ are
shown. It is found that $q$ decreases as the orbital angular momentum increases.

\begin{figure*}[htbp]
	\center{
	\subfigure[E=0.964295]{\label{Periodicorbits6a}
	\includegraphics[width=5.1cm]{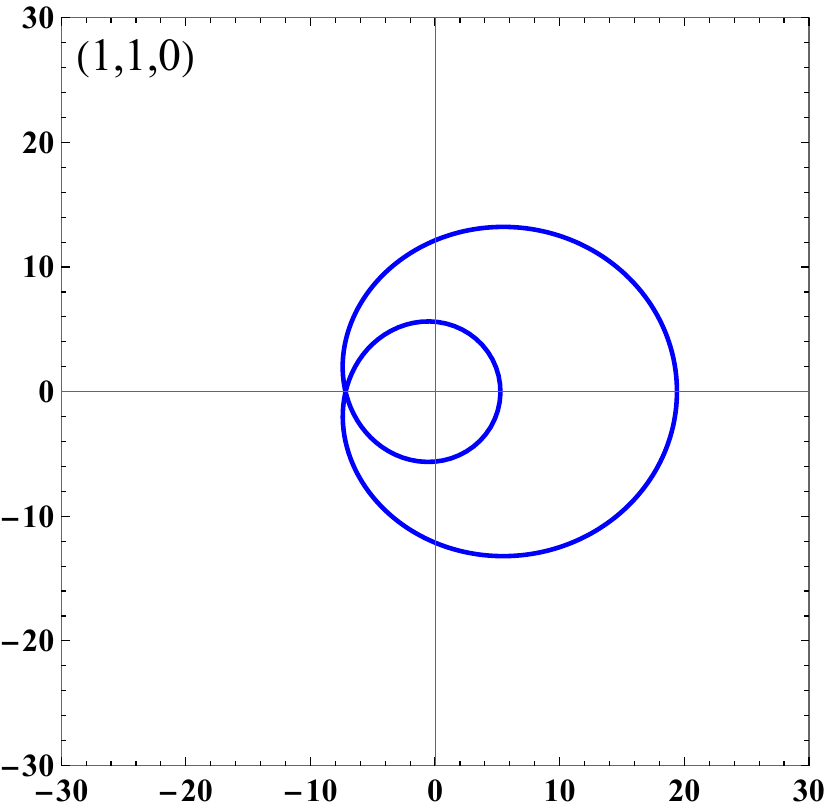}}
	\subfigure[E=0.967992]{\label{Periodicorbits6b}
	\includegraphics[width=5.1cm]{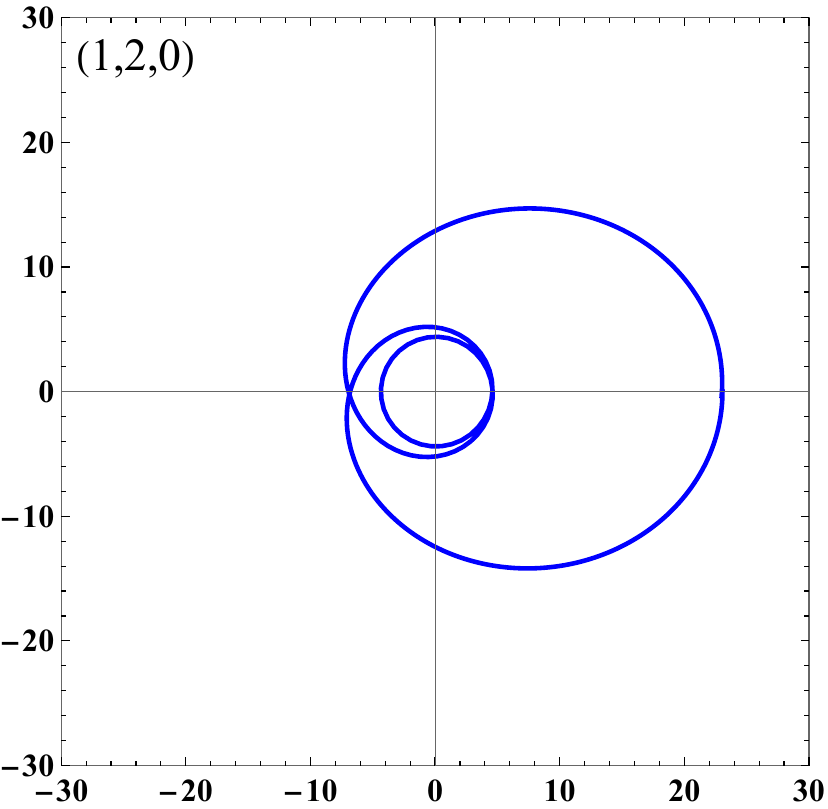}}
	\subfigure[E=0.968107]{\label{Periodicorbits6c}
	\includegraphics[width=5.1cm]{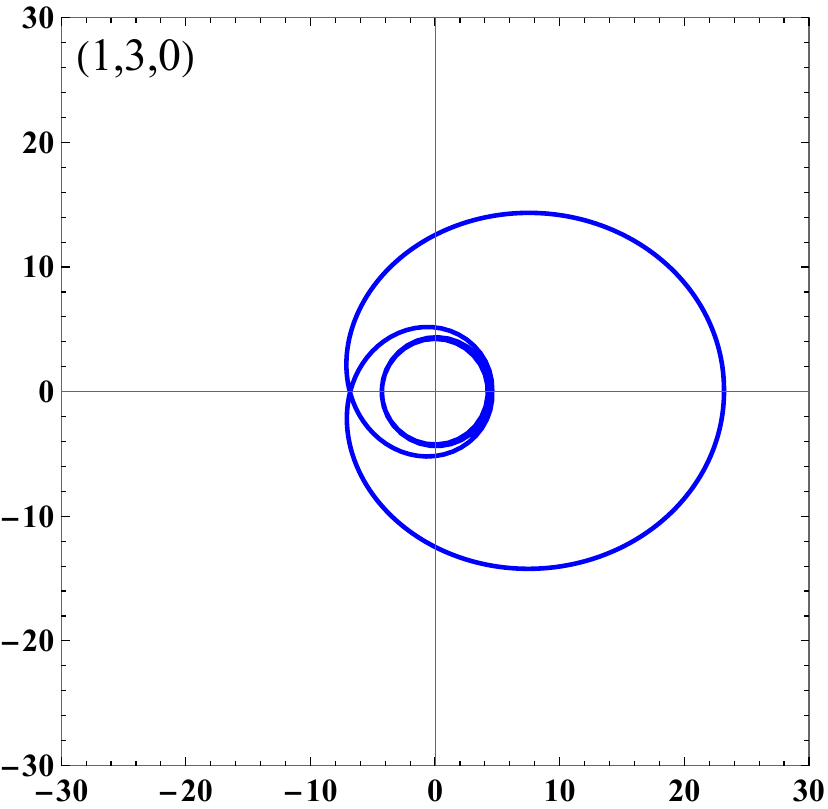}}\\
	\subfigure[E=0.967469]{\label{tPeriodicorbits6d}
	\includegraphics[width=5.1cm]{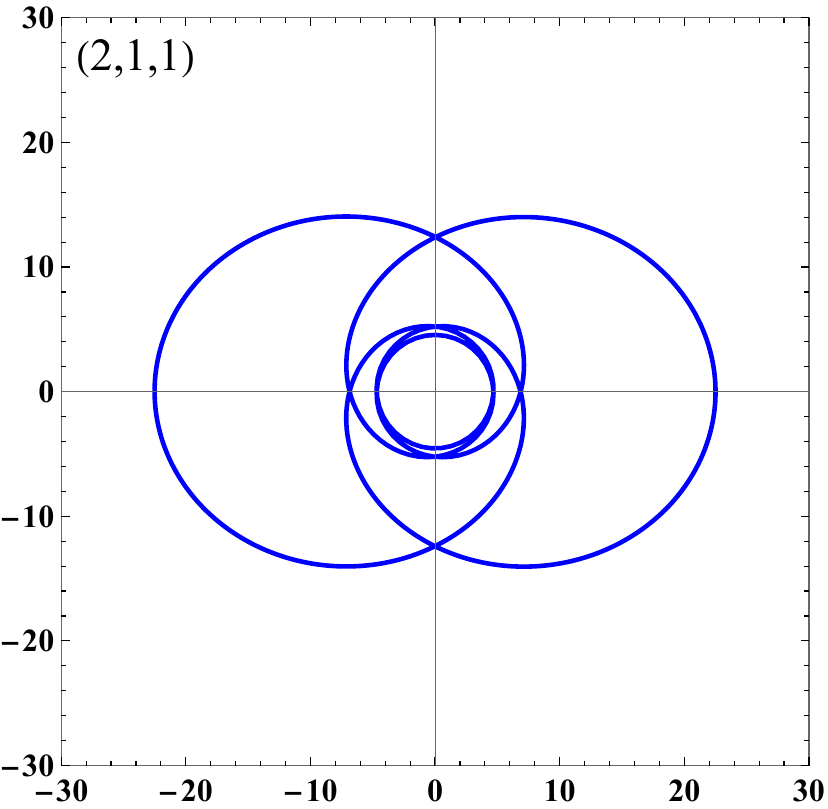}}
	\subfigure[E=0.968091]{\label{Periodicorbits6e}
	\includegraphics[width=5.1cm]{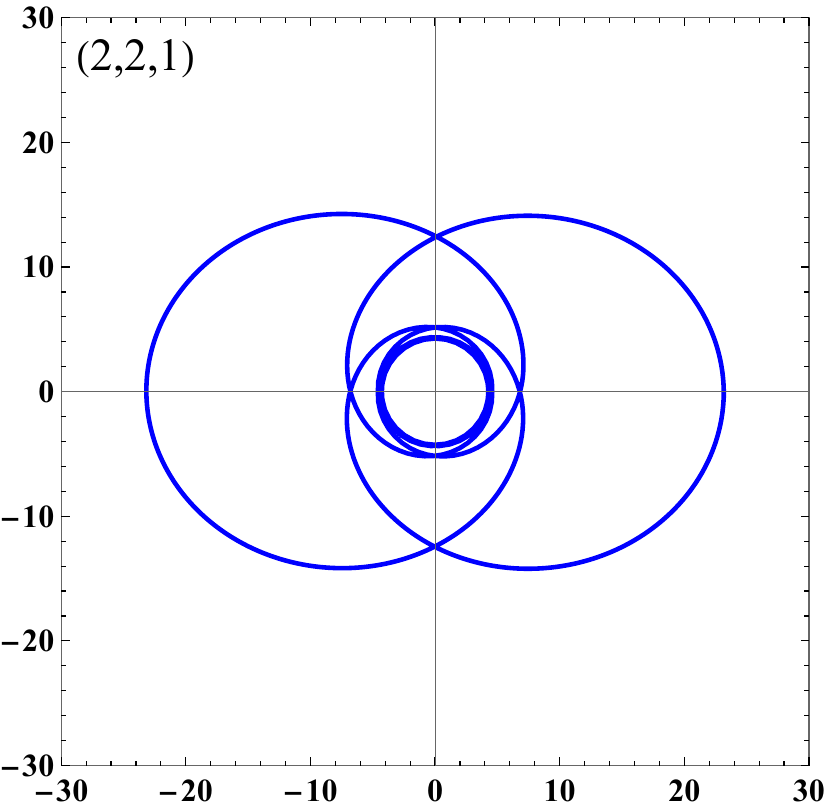}}
	\subfigure[E=0.968110]{\label{Periodicorbits6f}
	\includegraphics[width=5.1cm]{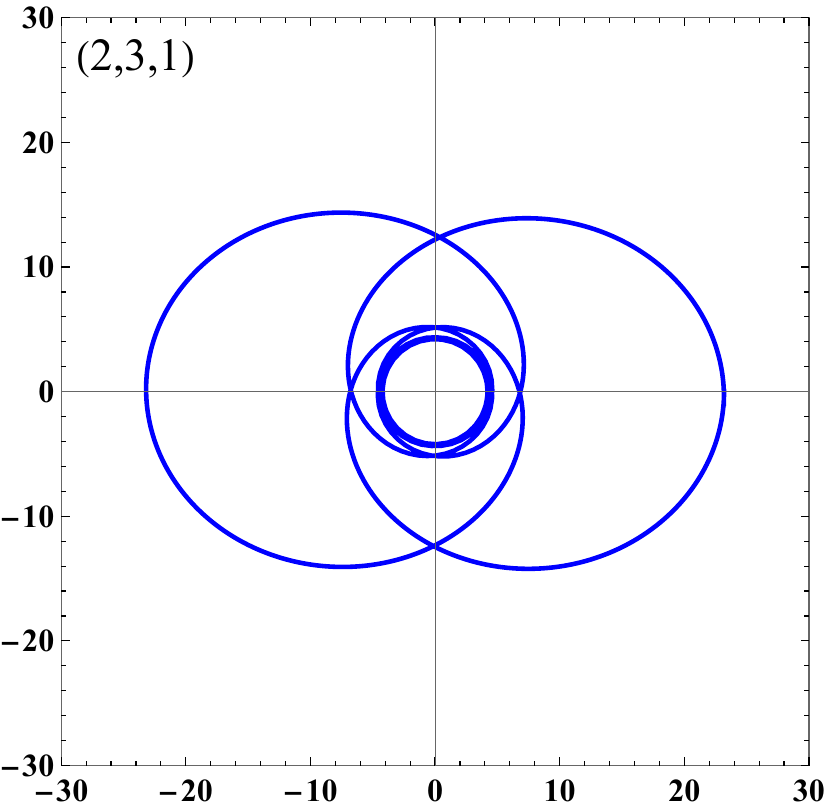}}
	\subfigure[E=0.967748]{\label{Periodicorbits6g}
	\includegraphics[width=5.1cm]{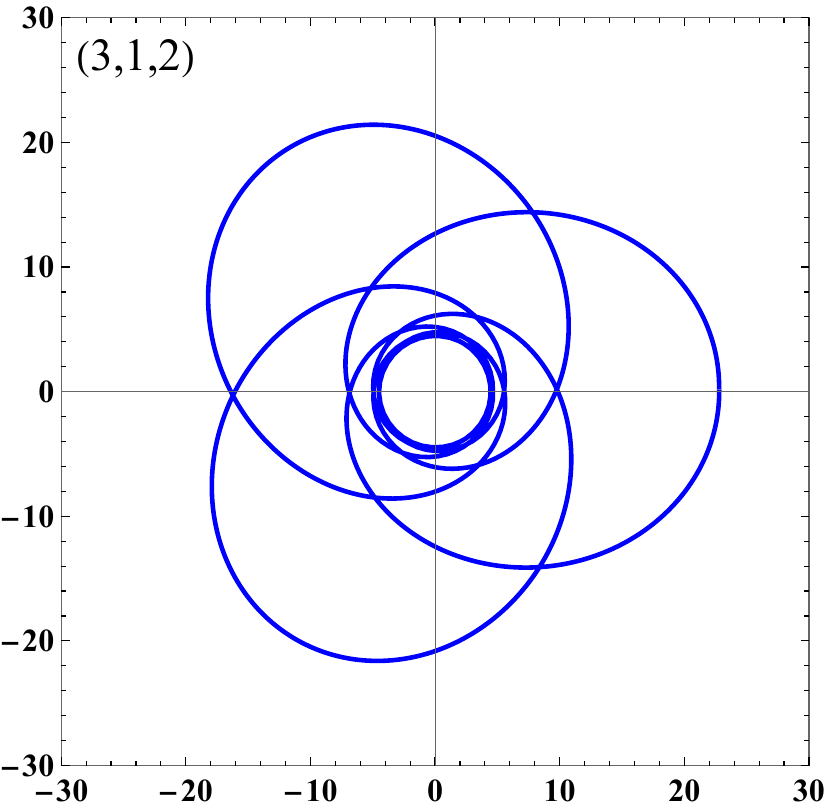}}
	\subfigure[E=0.968099]{\label{Periodicorbits6h}
	\includegraphics[width=5.1cm]{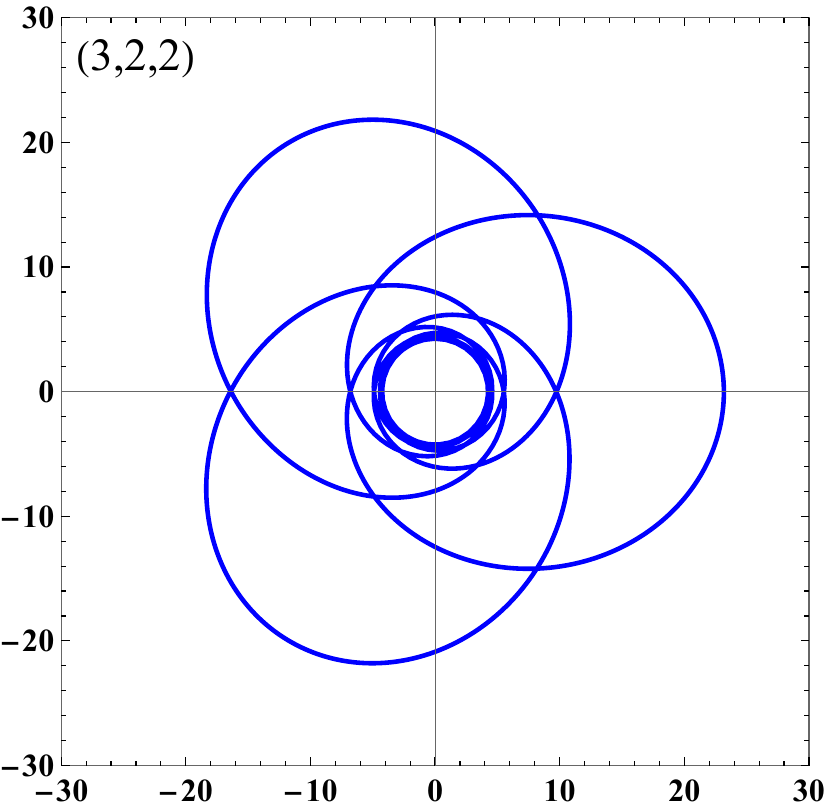}}
	\subfigure[E=0.968111]{\label{Periodicorbits6i}
	\includegraphics[width=5.1cm]{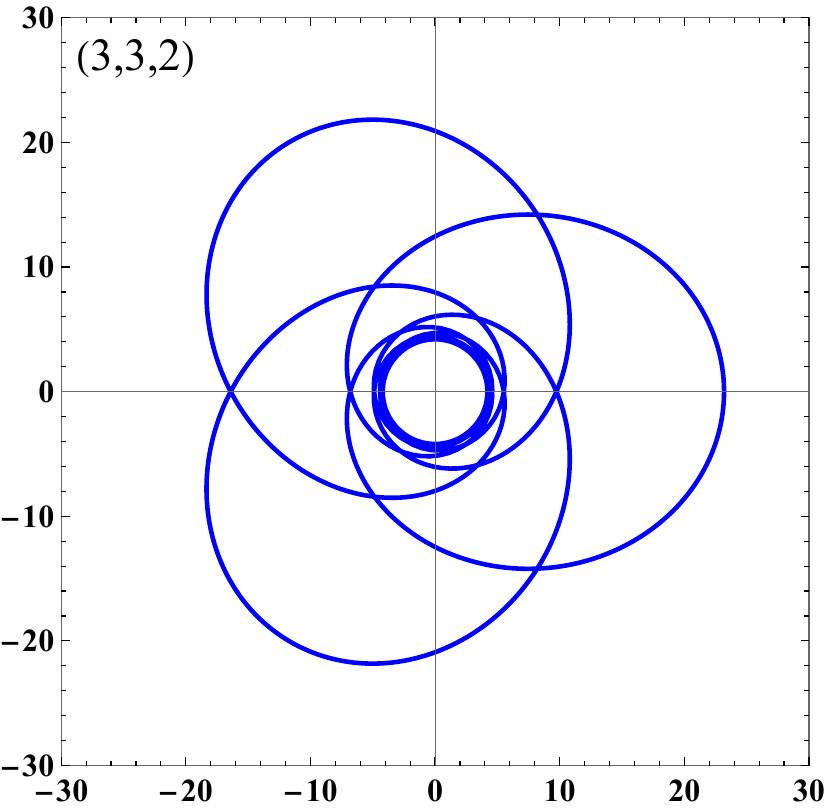}}\\
	\subfigure[E=0.958364]{\label{tPeriodicorbits6j}
	\includegraphics[width=5.1cm]{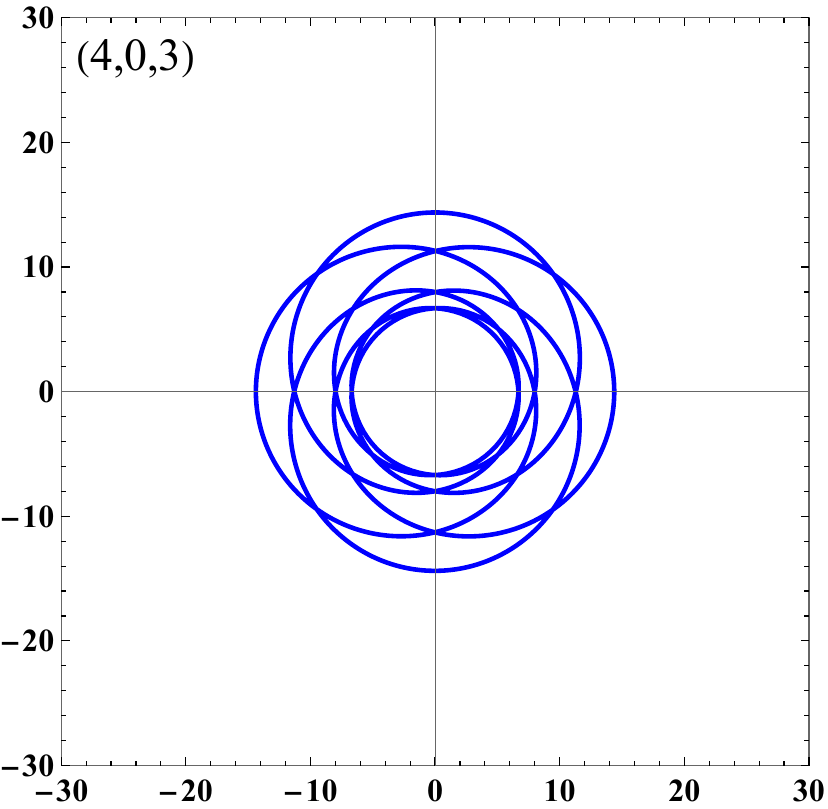}}
	\subfigure[E=0.967841]{\label{Periodicorbits6k}
	\includegraphics[width=5.1cm]{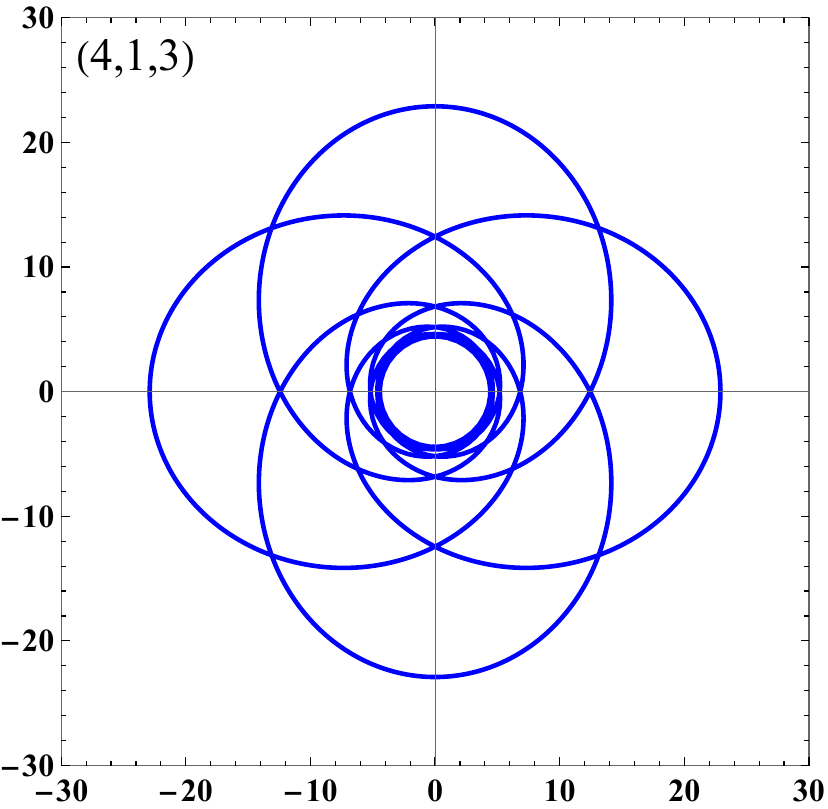}}
	\subfigure[E=0.968102]{\label{Periodicorbits6l}
	\includegraphics[width=5.1cm]{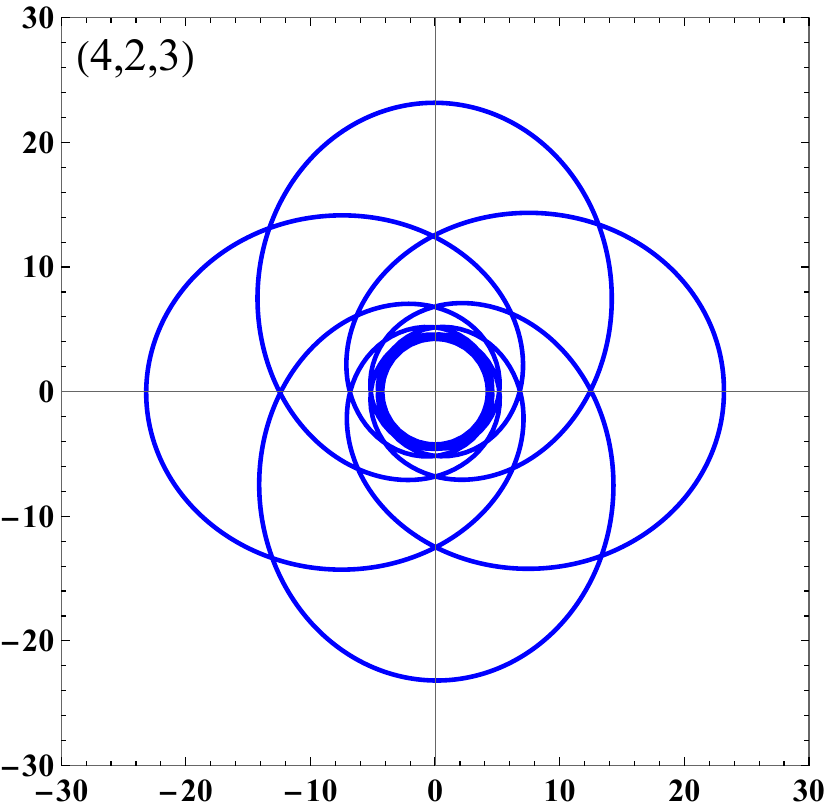}}
	}
\captionsetup{justification=raggedright}
	\caption{Periodic orbits characterized by different values of $(z, w, v)$ around the black hole without Cauchy horizon with $\ell/\mathcal{M}=1$ and $\epsilon=0.5$.}\label{E-trajectory}
\end{figure*}

\begin{figure*}[htbp]
	\center{
	\subfigure[$L/\mathcal{M}$=3.680690]{\label{Periodicorbits7a}
	\includegraphics[width=5.1cm]{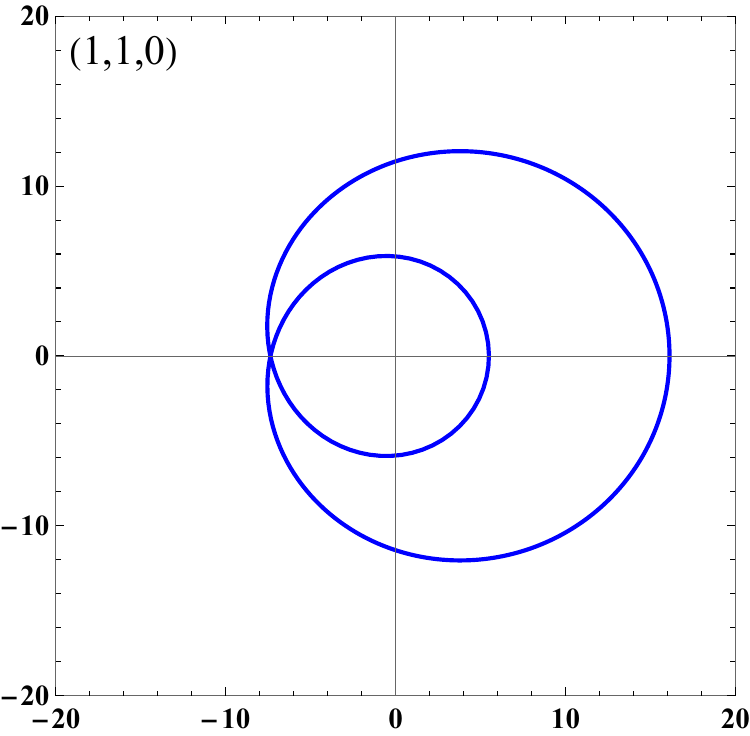}}
	\subfigure[$L/\mathcal{M}$=3.646054]{\label{Periodicorbits7b}
	\includegraphics[width=5.1cm]{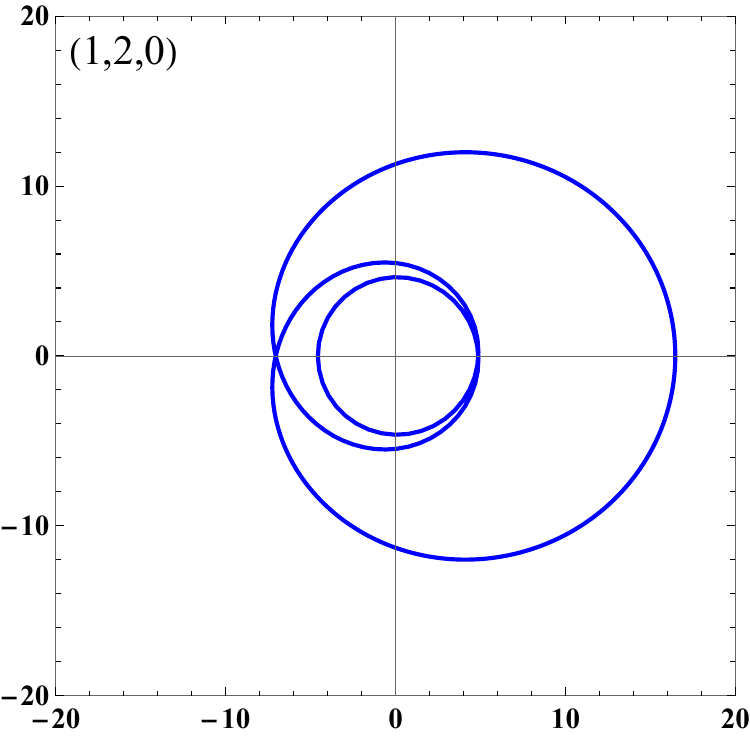}}
	\subfigure[$L/\mathcal{M}$=3.646050]{\label{Periodicorbits7c}
	\includegraphics[width=5.1cm]{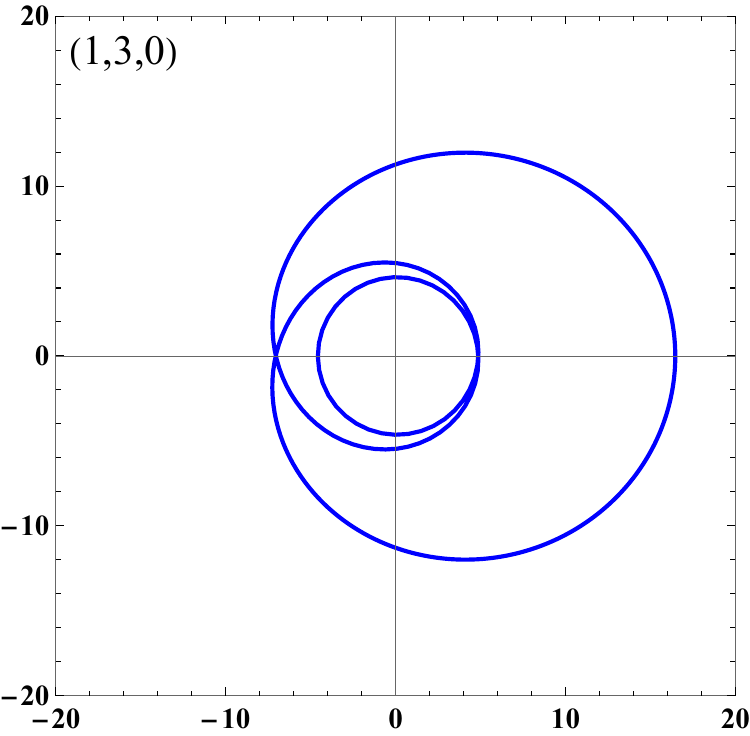}}\\
	\subfigure[$L/\mathcal{M}$=3.651467]{\label{tPeriodicorbits7d}
	\includegraphics[width=5.1cm]{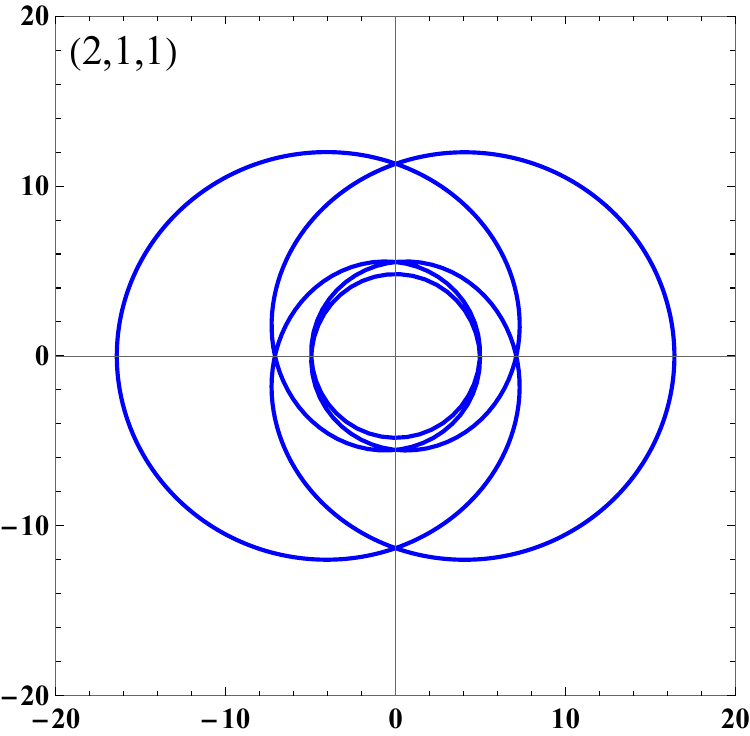}}
	\subfigure[$L/\mathcal{M}$=3.644985]{\label{Periodicorbits7e}
	\includegraphics[width=5.1cm]{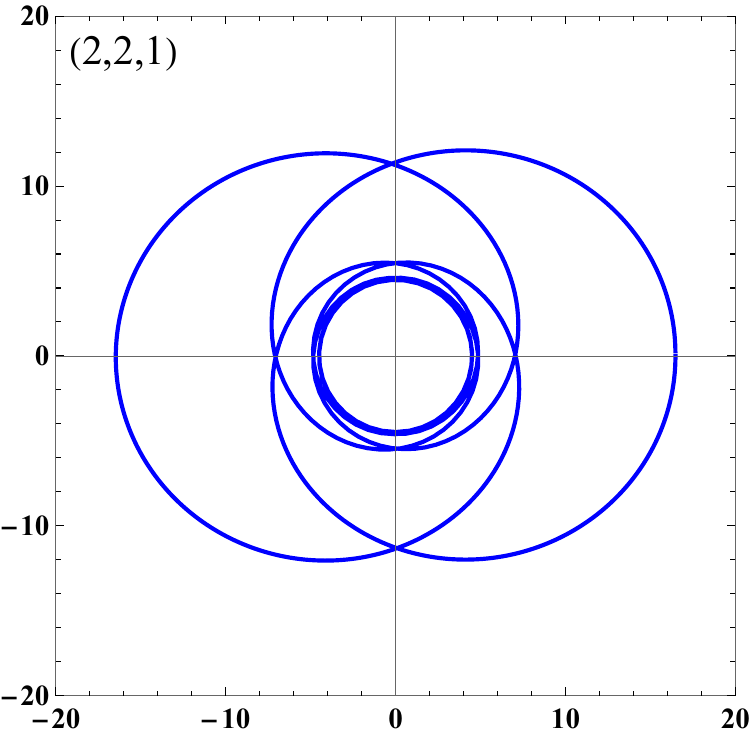}}
	\subfigure[$L/\mathcal{M}$=3.644985]{\label{Periodicorbits7f}
	\includegraphics[width=5.1cm]{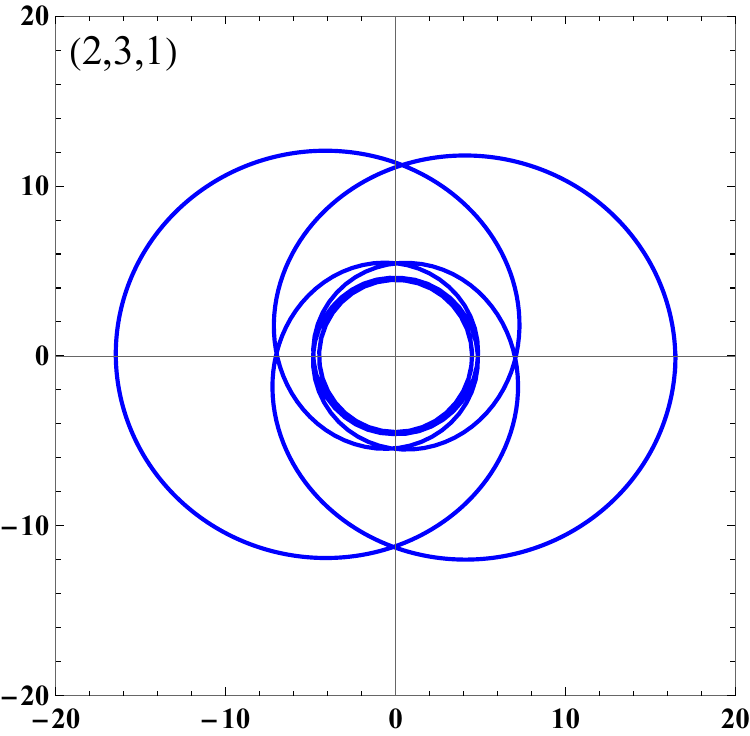}}
	\subfigure[$L/\mathcal{M}$=3.648655]{\label{Periodicorbits7g}
	\includegraphics[width=5.1cm]{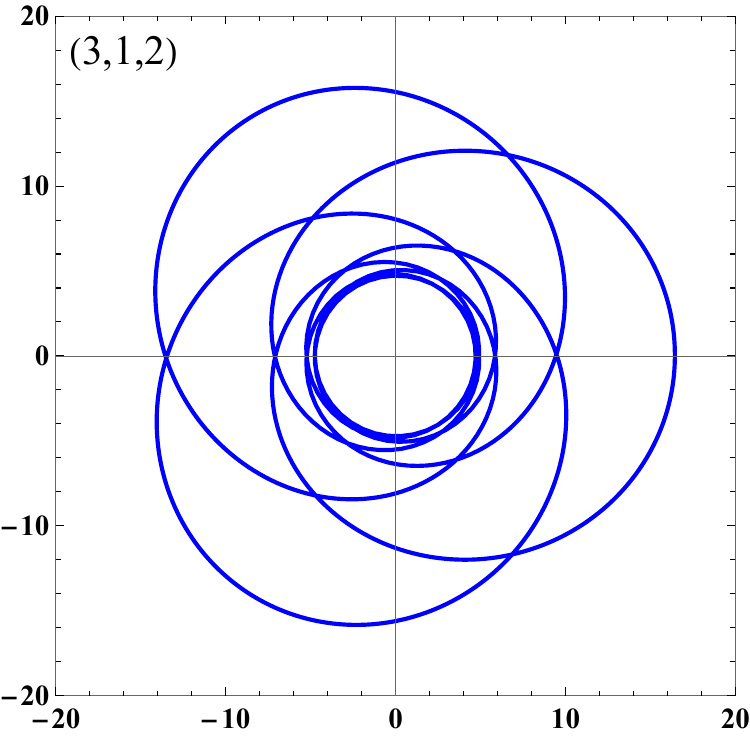}}
	\subfigure[$L/\mathcal{M}$=3.645175]{\label{Periodicorbits7h}
	\includegraphics[width=5.1cm]{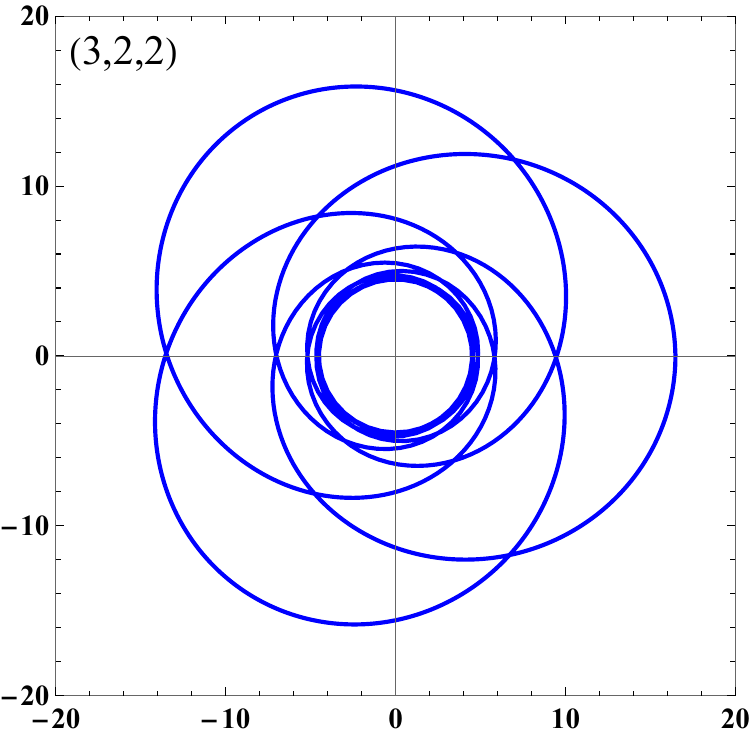}}
	\subfigure[$L/\mathcal{M}$=3.644878]{\label{Periodicorbits7i}
	\includegraphics[width=5.1cm]{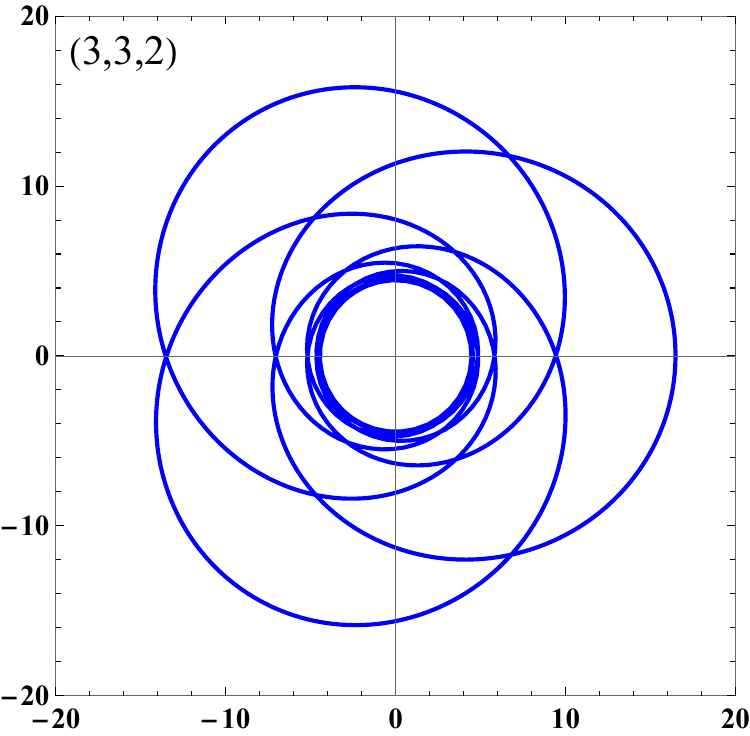}}\\
	\subfigure[$L/\mathcal{M}$=3.731895]{\label{tPeriodicorbits7j}
	\includegraphics[width=5.1cm]{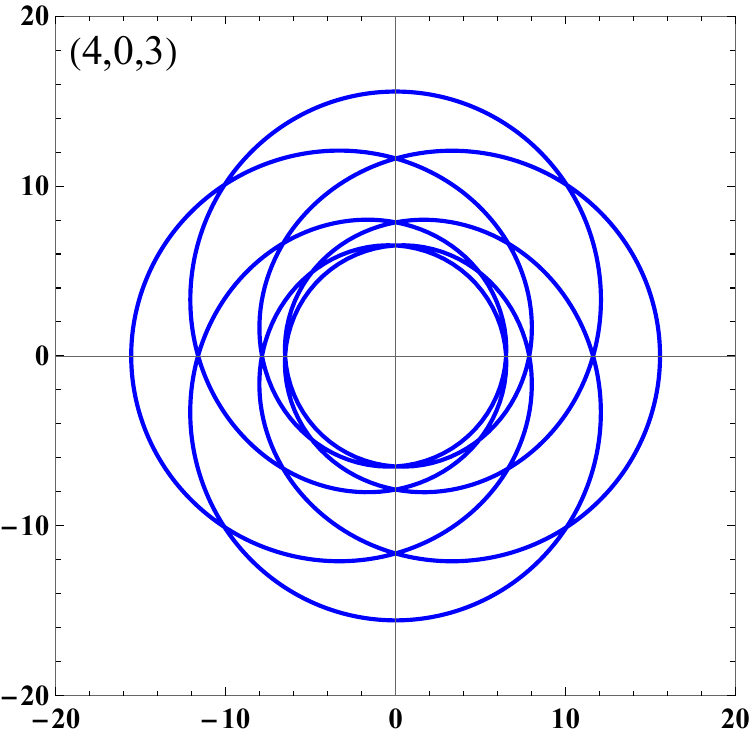}}
	\subfigure[$L/\mathcal{M}$=3.647710]{\label{Periodicorbits7k}
	\includegraphics[width=5.1cm]{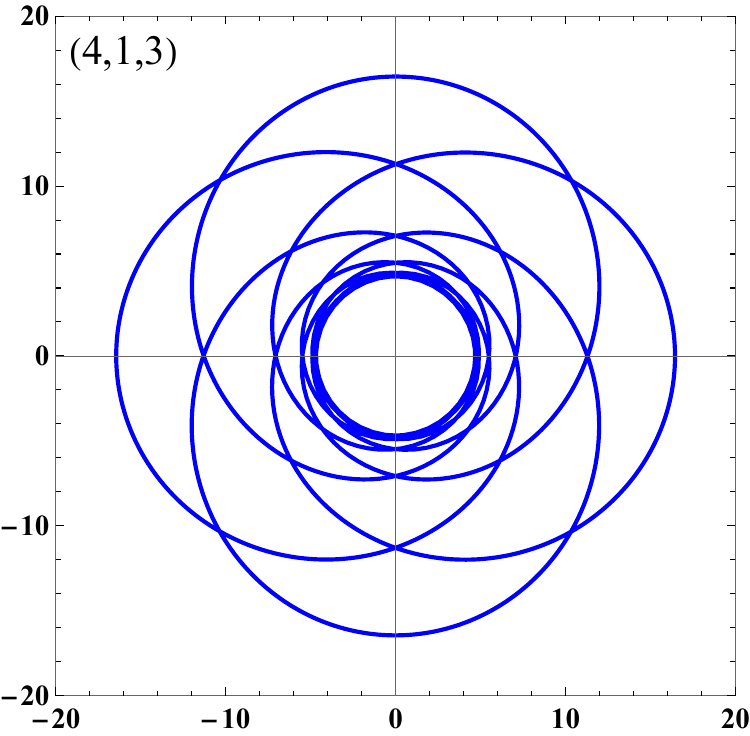}}
	\subfigure[$L/\mathcal{M}$=3.645312]{\label{Periodicorbits7l}
	\includegraphics[width=5.1cm]{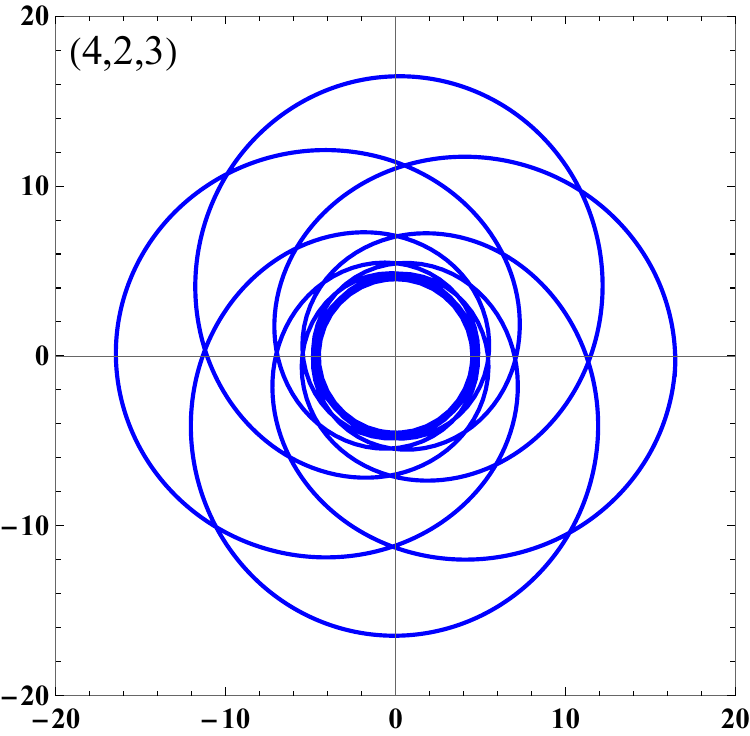}}
	}
\captionsetup{justification=raggedright}
	\caption{Periodic orbits characterized by different values of $(z, w, v)$ around the black hole without Cauchy horizon with $\ell/\mathcal{M}=1$ and $E=0.96$.}\label{L-trajectory}
\end{figure*}

In Fig.~\ref{E-trajectory}, we present the trajectories of periodic orbits with different $(z,w,v)$ in the $r-\phi$ plane for fixed hair parameter $\ell/\mathcal{M} =1$ and $\epsilon=0.5$ ($L/\mathcal{M}$=3.71792), in the background of a black hole without Cauchy horizon. The horizontal and vertical coordinates represent $r cos\phi$ and $r sin\phi$, respectively. The energy  $E$  and $q$ corresponding to the particle and orbit are marked in the subfigures. In Fig.~\ref{L-trajectory}, we show the trajectories of periodic orbits with different $(z,w,v)$  for a fixed black hole hair parameter $\ell/\mathcal{M}=1$ and particle energy $E=0.96$. By comparing Fig.~\ref{E-trajectory} and Fig.~\ref{L-trajectory}, it is evident that for the same $q$, the trajectory patterns of the periodic orbits are identical. The figure clearly shows that the integer $z$ describes the number of leaf-like patterns in the trajectory. As $z$ increases, the shape of the leaves becomes larger and the orbits become more complex.

Moreover, after fixing $\epsilon=0.5$, the particle energies corresponding to different $q$-periodic orbits for varying hair parameter $\ell/{\mathcal{M}}$ are listed in Table. \ref{tab:energylist}. The orbital angular momentum corresponding to $\epsilon=0.5$ is shown in the second column. It can be observed that as the black hole hair parameter increases, both the orbital angular momentum and particle energy increase.  Their maximum values will be reached as the system approaches the Schwarzschild solution.

\begin{table*}[htbp]
	\caption{The particle energy $E$ of the periodic orbits characterized by different $(z, w, v)$ configurations with $\epsilon = 0.5$. \label{tab:energylist}}
	\begin{center}
	\vspace{-0.2cm}
	\begin{tabular}{cccccccccc}
	\toprule[0.5pt]\toprule[0.5pt]
	$\ell/{\mathcal{M}}$  & $L/{\mathcal{M}}$ &  $E_{(1,1,0)}$    & $E_{(1,2,0)}$  & $E_{(2,1,1)}$  &  $E_{(2,2,1)}$  & $E_{(3,1,2)}$  & $E_{(3,2,2)}$  & $E_{(4,1,3)}$ & $E_{(4,2,3)}$ \\
	\midrule[0.5pt]
	1   & 3.71792  &0.964295   & 0.967992  & 0.967469  & 0.968091 & 0.967750 &  0.968099 & 0.967841 & 0.968102 \\
	1.05& 3.722201 &0.964575   & 0.968103  & 0.967626  & 0.968184 & 0.967883 &  0.968191 & 0.967965 & 0.9681938\\
	1.1 & 3.725260 &0.964857   & 0.968183  & 0.967733  & 0.968252 & 0.967977 &  0.968258 & 0.968054 & 0.968265 \\
	1.2 & 3.729044 &0.965159   & 0.968279  & 0.967889  & 0.968340 & 0.968101 &  0.968345 & 0.968167 & 0.968347  \\
	1.99& 3.732051 &0.965427   & 0.968382  & 0.968025  & 0.968434 & 0.968225 & 0.968439  & 0.968285 & 0.968439  \\
	\bottomrule[0.5pt] \bottomrule[0.5pt]
	\end{tabular}
	\end{center}
	\end{table*}

In Table \ref{tab:angmomlisy}, we also list the orbital angular momentum $L/{\mathcal{M}}$ of the particle corresponding to different periodic orbits $(z,w,v)$ at a fixed particle energy $E=0.96$ for various black hole hair parameter $\ell/{\mathcal{M}}$. Similar to the results shown in Table. \ref{tab:energylist}, as the black hole hair parameter increases, the orbital angular momentum of the corresponding particle also increases.

\begin{table*}
 \caption{The angular momentum $L/{\mathcal{M}}$ of the periodic orbits characterized by different $(z, w, v)$ configurations with energy $E = 0.96$.} \label{tab:angmomlisy}
	\begin{center}
	\vspace{-0.2cm}
	\begin{tabular}{ccccccccc}
	\toprule[0.5pt]\toprule[0.5pt]
	$\ell/{\mathcal{M}}$    &  $L/{\mathcal{M}}_{(1,1,0)}$    & $L/{\mathcal{M}}_{(1,2,0)}$  & $L/{\mathcal{M}_{(2,1,1)}}$  &  $L/{\mathcal{M}}_{(2,2,1)}$  & $L/{\mathcal{M}}_{(3,1,2)}$  & $L/{\mathcal{M}}_{(3,2,2)}$  & $L/{\mathcal{M}}_{(4,1,3)}$ & $L/{\mathcal{M}}_{(4,2,3)}$ \\
	\midrule[0.5pt]
		  1         & 3.680690 &3.646054&3.651467&3.644990  & 3.648655  & 3.645175 & 3.647710 & 3.645312 \\
		 1.05       & 3.681680 &3.648460&3.653510&3.647501  & 3.650850  & 3.647661 & 3.649959 & 3.647789 \\
		 1.1        & 3.682471 &3.650099&3.654953&3.649259  & 3.652396  & 3.649406 & 3.651547 & 3.649524 \\
		1.2         & 3.683335 &3.652135&3.656614&3.651331  & 3.654198  & 3.654153 & 3.653435 & 3.651233 \\
		1.99        & 3.683588 &3.653406&3.657596&3.652701  & 3.655315  & 3.652815 & 3.654621 & 3.652615 \\
	\bottomrule[0.5pt] \bottomrule[0.5pt]
	\end{tabular}
	\end{center}
	\end{table*}

\section{Gravitational wave radiation from periodic orbits}
\label{GWandpo}

In the study of black hole orbital systems, we typically treat the supermassive black hole without the Cauchy horizon as the central object, with a smaller mass object acting as a test particle orbiting the central body. As the particle approaches the center black hole, it experiences an increasingly stronger gravitational field, causing changes in the curvature of spacetime and generating GWs. These GWs take away the angular momentum and energy of the particles, leading to the particles gradually spiraling inward and moving toward the black hole. In such a system, compared to the total energy, the energy and angular momentum loss due to the motion of the smaller mass object is negligible over a few periods, which allows the application of the adiabatic approximation \cite{Grossman:2011im, Sundararajan:2007jg, Hughes:1999bq, Hughes:2005qb, Zi:2023qfk, Zi:2024dpi}. The adiabatic approximation assumes that over the course of one orbital period, the system's energy and angular momentum remain nearly constant, allowing us to neglect the effect of gravitational radiation on the motion of the smaller object. Under this assumption, the motion of the test particle can be described using the geodesic equations, and the resulting GW forms can be derived using the quadrupole formula by the ``Kludge'' method \cite{Babak:2006uv}.

As the smaller mass object moves in a periodic orbit around the supermassive black hole, it traces a unique bound trajectory, oscillating between the apoapsis and periapsis of the orbit. Notably, as the particle spirals inward, it experiences long-duration, high-frequency oscillations near the event horizon, generating rich gravitational wave signals. These waves carry crucial information about the physics near the black hole's event horizon, making them one of the most promising tools for probing the nature of black holes. Therefore, the GWs emitted by EMRIs provide a direct link between the motion of the test particle and the underlying physics of the black hole, offering an invaluable means for studying both the dynamics of compact objects and the fundamental characteristics of black holes.

The Kludge GW forms are derived by calculating the trajectory of the particle in the Boyer-Lindquist coordinate system, which is then mapped to the polar coordinates of a flat-space spherical polar coordinates. Finally, the GW multipole moments are obtained from the particle's trajectory in these polar coordinates \cite{Babak:2006uv}. Firstly, we treat the Boyer-Lindquist coordinates as a hypothetical spherical coordinate system and project the trajectory of the small object $(r, \theta, \phi)$ onto the
Cartesian coordinates $(x, y, z)$, as shown below
\begin{equation}\label{cartesian coor}
x=r \sin \theta \cos \phi, \quad y=r \sin \theta \sin \phi, \quad z=r \cos \theta .
\end{equation}
After constructing the particle's orbit in our pseudo-flat space, we then apply the wave generation formula in flat space. In the weak-field linear approximation (i,e.  $g_{\mu\nu}=\eta_{\mu\nu}+h_{\mu\nu},~\bar{h}^{\mu\nu}\equiv h^{\mu\nu}-(1/2)h\eta^{\mu\nu}$, where $g_{\mu\nu}$, $\eta_{\mu\nu}$, $h_{\mu\nu}$ are spacetime metric, flat metric, small perturbations, respectively, and $|h_{\mu\nu}|\ll 1$, $h=\eta^{\mu\nu}h_{\mu\nu}$), and under the Lorentz gauge condition $\bar{h}^{\mu\alpha}_{,\alpha}=0$, the linearized Einstein field equation can be expressed as follows
\begin{equation}\label{linearizedEFE}
\square\bar{h}^{\mu\nu}=-16\pi\mathcal{T}^{\mu\nu},
\end{equation}
where $\square$ represents the standard flat-space wave operator, and the effective energy-momentum tensor $\mathcal{T}^{\mu\nu}$ satisfies $\mathcal{T}^{\mu\nu}_{,\nu}=0$. In the coordinate system centered on the black hole, one solution to the above equation is the retarded potential solution
\begin{equation}\label{retardedpotential}
h_{\mu\nu}(\mathbf{x},t)=4G \int d^3\mathbf{x}^{\prime} \frac{\mathcal{T}^{\mu\nu}(\mathbf{x}^{\prime},t-|\mathbf{x}-\mathbf{x}^{\prime}|)}{|\mathbf{x}-\mathbf{x}^{\prime}|},
\end{equation}
where $\mathbf{x}^{\prime}$ and $\mathbf{x}$ denote the positions of the observer and source, respectively. This solution represents the gravitational radiation emitted by the source $\mathcal{T}^{\mu\nu}$. The coordinate  $\mathbf{x}^{\prime}$ of the observer position serves as an integration variable that spans all possible locations in space where the effective energy-momentum tensor $\mathcal{T}^{\mu\nu}$ is non-zero. If the source motion is only negligibly influenced by gravity, $\mathcal{T}^{\mu\nu}$ can be taken to be equal to the energy-momentum tensor $T^{\mu\nu}$ of the matter source \cite{Babak:2006uv}. In the slow motion limit, the Press formula reduces to the usual quadrupole formula \cite{Babak:2006uv}
\begin{equation}\label{quadrupoleformula}
\bar{h}^{i j}(t, \mathbf{x})=\frac{2}{r}\left[\ddot{I}^{i j}\left(t^{\prime}\right)\right]_{t^{\prime}=t-r},
\end{equation}
where
\begin{equation}\label{quadrupolemoment}
I^{i j}=\int  x^i x^j T^{t t}\left(t, x^i\right) d^3x,
\end{equation}
is the source mass quadrupole moment. Here, the $T^{t t}$ component of the stress-energy tensor for the small object with trajectory $Z^i(t)$ is \cite{Thorne:1980ru}:
\begin{equation}\label{stress-energy tensor}
T^{t t}\left(t, x^i\right)=m \delta^3\left(x^i-Z^i(t)\right) .
\end{equation}
By substituting \eqref{quadrupolemoment} into \eqref{quadrupoleformula}, we obtain the GW quadrupole formula under the slow-rotation approximation
\begin{equation}\label{GW}
h_{ij}=\frac{2}{D_{\mathrm{L}}}\frac{d^2I_{ij}}{dt^2}=\frac{2m}{D_{\mathrm{L}}}(a_ix_j+a_jx_i+2v_iv_j),
\end{equation}
Here, $D_{\mathrm{L}}$ denotes the luminosity distance from the EMRIs to the detector, while $v_i$  and  $a_i$ represent the spatial velocity and acceleration of the small object, respectively.

Finally, one can project the above GW onto the detector-adapted coordinate system \cite{Poisson:2014}. Then the corresponding plus $h_+$ and cross $h_\times$ polarizations of the GW are given by
\begin{equation}\label{polarizations}
\begin{aligned}
		h_+&=(h_{\Theta \Theta}-h_{\Phi\Phi})/2,\\
		h_\times&=h_{\Theta\Phi},
\end{aligned}
\end{equation}
where the components $h_{\Theta \Theta}$, $h_{\Phi \Theta}$ and $h_{\Phi \Phi}$ are given by \cite{Babak:2006uv}:
\begin{equation}\label{component}
\begin{aligned}
h_{\Theta \Theta}=&\cos ^2 \Theta\left[h_{x x} \cos ^2 \Phi+h_{x y} \sin 2 \Phi+h_{y y} \sin ^2 \Phi\right]\\
                  &+h_{z z} \sin ^2 \Theta-\sin 2 \Theta\left[h_{x z} \cos \Phi+h_{y z} \sin \Phi\right], \\
h_{\Phi \Theta}=&\cos \Theta\left[-\frac{1}{2} h_{x x} \sin 2 \Phi+h_{x y} \cos 2 \Phi+\frac{1}{2} h_{y y} \sin 2 \Phi\right]\\
                &+\sin \Theta\left[h_{x z} \sin \Phi-h_{y z} \cos \Phi\right], \\
h_{\Phi \Phi}=&h_{x x} \sin ^2 \Phi-h_{x y} \sin 2 \Phi+h_{y y} \cos ^2 \Phi .
\end{aligned}
\end{equation}

To demonstrate the GW forms for different periodic orbits and how the hair parameter $\ell$ affects them, we consider an EMRI system consisting of a smaller object with mass $m = 10 M_\odot$ and a supermassive black hole with mass $\mathcal{M} = 10^7 M_\odot$, where $M_\odot$ is the solar mass. For simplicity, we set the inclination angle $\Theta=\pi/4$, latitude angle $\Phi=\pi/4$, and the luminosity distance $D_L=200~\text{Mpc}$.

\begin{figure*}[htbp]
    \centering
    \begin{minipage}{0.3\textwidth}
        \centering
        \includegraphics[width=\linewidth]{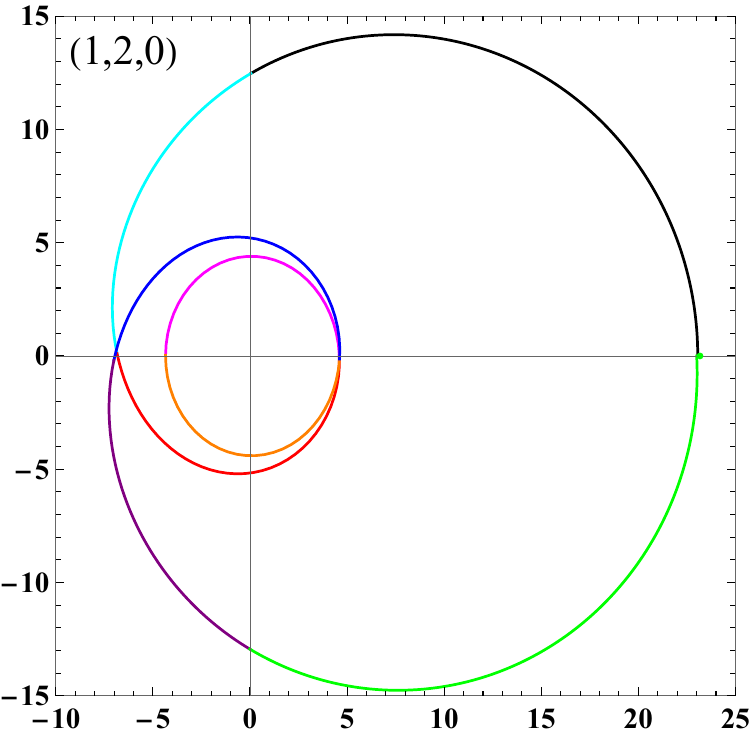}
        (a) Periodic orbit.
    \end{minipage}
    \hfill
    \begin{minipage}{0.65\textwidth}
        \centering
        \includegraphics[width=\linewidth]{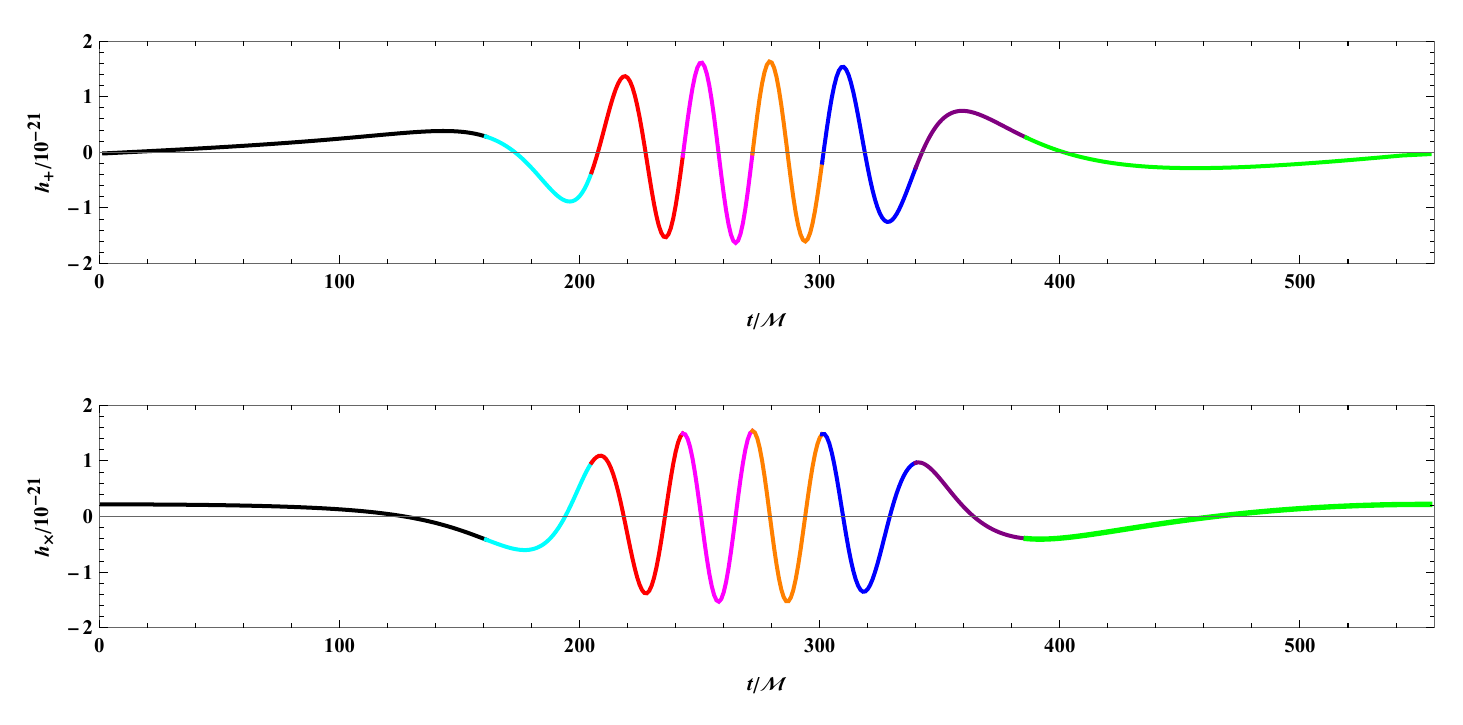}
		(b) Gravitational waveforms.
    \end{minipage}
  \captionsetup{justification=raggedright}
    \caption{Gravitational waveform corresponding to the periodic orbit with $q=(1,2,0)$. (a) Periodic orbit. (b) Gravitational waveform. Different segments are marked in different colors.}
    \label{fig:POGW}
\end{figure*}

In Fig.~\ref{fig:POGW}, we use different colors to highlight the correspondence between the GW form and the periodic orbit, enabling a detailed analysis of the relationship between the periodic orbit and the resulting gravitational radiation. First, the GW form clearly captures the zoom-whirl behavior of the periodic orbit over a complete cycle, reflecting the zooming and whirling motions of the small object trajectory. Second, as shown in Fig.~\ref{fig:POGW}, the waveform amplitude peaks when the orbit approaches perihelion and decreases as the orbit moves farther from perihelion. From this, we can conclude that the smooth portion of the waveform associated with the periodic orbit corresponds to the zoom phase, during which the small object traverses a highly elliptical orbit far from the black hole. In contrast, the sharply oscillating portion of the waveform corresponds to the whirl phase, occurring when the small object approaches the black hole and undergoes circular whirl motion. In the early stages, when the particle is far from the black hole and the gravitational field is relatively weak, the waveform remains smooth and less pronounced. As the particle approaches the black hole, its trajectory becomes more distorted, the gravitational field strengthens, and both the amplitude and frequency of the GWs increase significantly. In the final stages, as the particle nears the event horizon, the GW signal reaches its peak, characterized by a sharp frequency sweep and a dramatic increase in amplitude, and the oscillations become more pronounced. As the particle accelerates and its trajectory becomes increasingly distorted, the amplitude and frequency of the gravitational waves gradually increase. Consequently, GWs carry crucial information about the particle trajectory, mass distribution, and other properties, making it possible to study the physical characteristics of black holes and their surroundings through waveform analysis.

\begin{figure*}[htbp]
	\center{
	\subfigure[$\ell=1 \mathcal{M}$]{\label{hpc8a}
	\includegraphics[width=16.6cm]{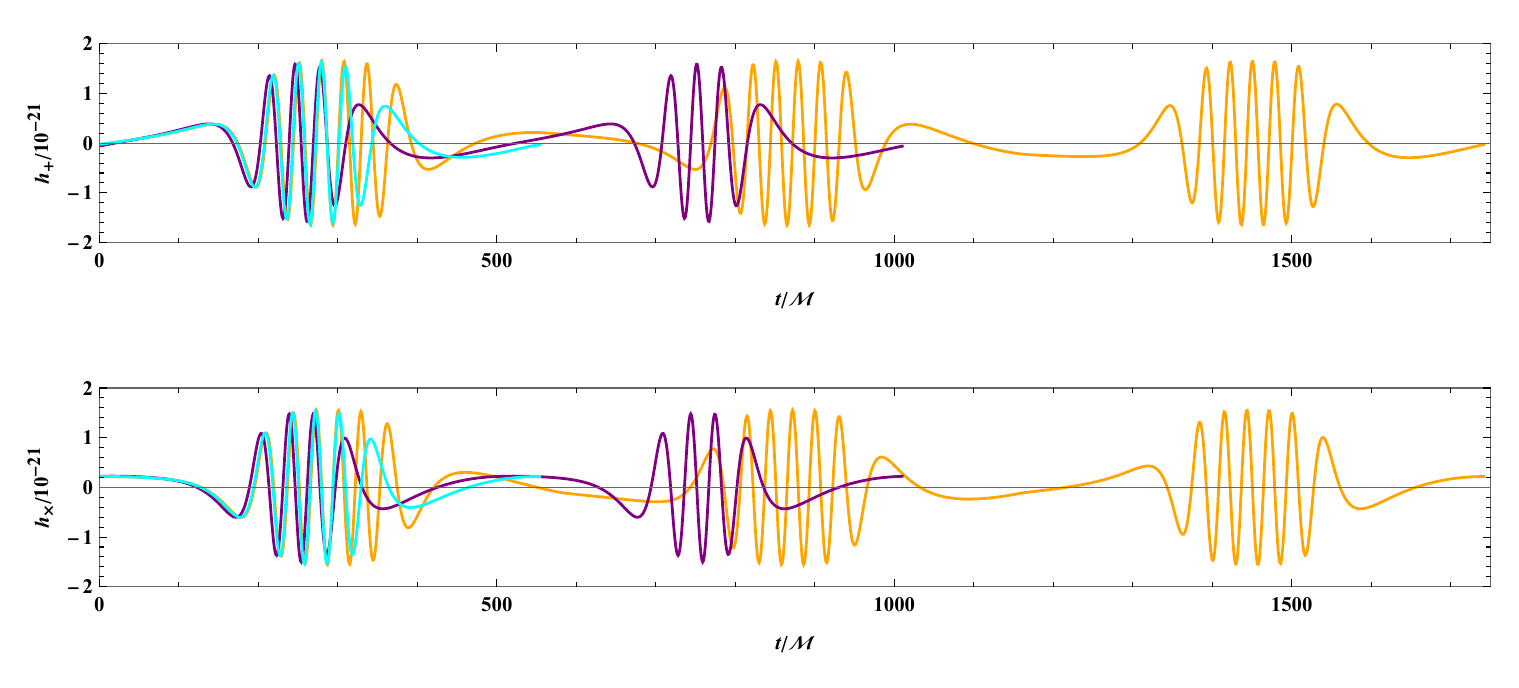}}\\
	\subfigure[$\ell=1.99 \mathcal{M}$]{\label{hpc8b}
	\includegraphics[width=16.6cm]{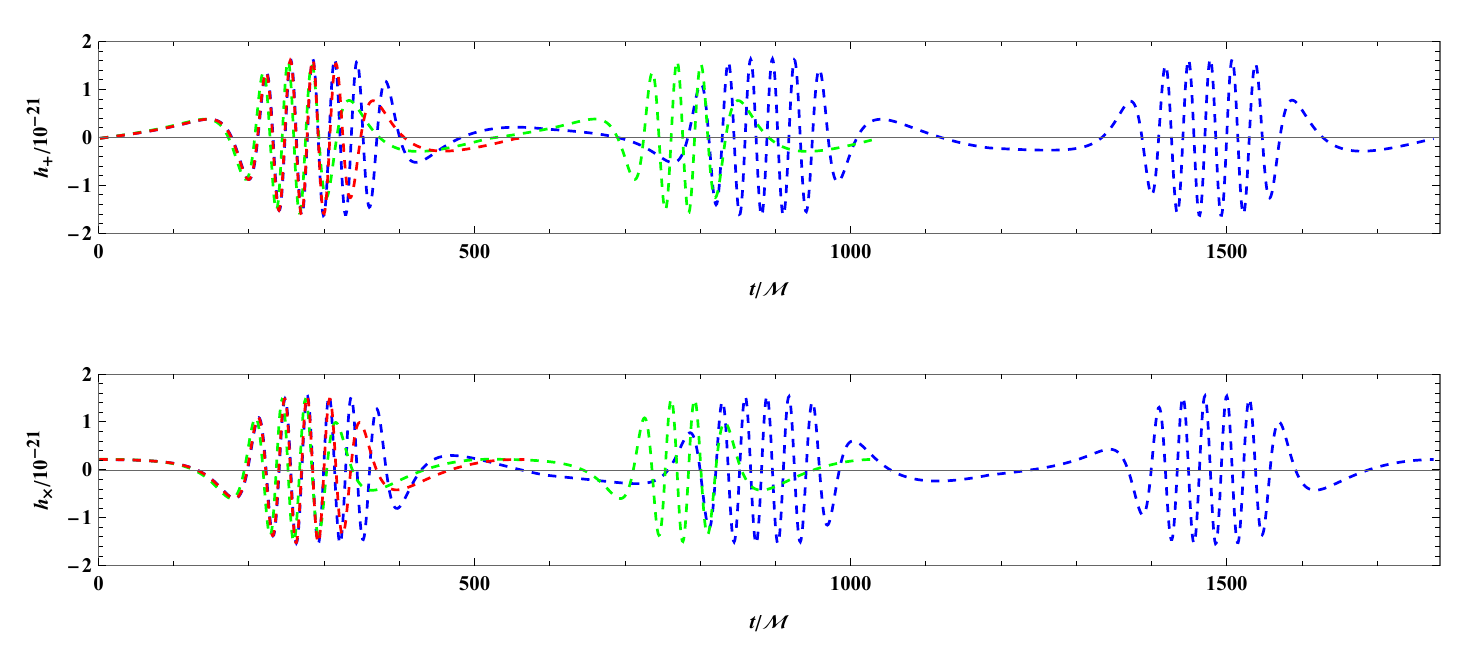}}\\
	}
\captionsetup{justification=raggedright}
\caption{The GW waveforms produced  by different periodic orbits for different $\ell$ with $\epsilon=0.5$. Different $q = q (z, w, v)$ are denoted with different colors. $(z, w, v) = (1,1, 0)$ is shown in cyan and red, $(z, w, v) = (2,1,1)$ in purple and green, $(z, w, v) = (3,2,2)$ in orange and blue. The solid curves indicate $\ell/\mathcal{M}=1$, and the dashed curves indicate $\ell/\mathcal{M}=1.99$, which is very close to the Schwarzschild case. (a) $\ell=1 M$. (b) $\ell=1.99 M$.}
\label{fig:waveforms1}
\end{figure*}

On the other hand, Fig.~\ref{fig:waveforms1} shows the GW forms corresponding to different $(z, w, v)$ periodic orbits. It is observed that the number of quiet phases in the waveform corresponds to the number of ``leaves" of the orbit, while the pronounced oscillations align with the number of whirls in the orbit. Compared to the Schwarzschild case $(\ell/\mathcal{M}= 1.99)$, the parameter $\ell/\mathcal{M}$ primarily affects the phase of the GW signal, with minimal (almost negligible) impact on the amplitude. These features suggest that the GW signals emitted by periodic orbits capture the essential characteristics of the zoom and whirl phases, which may help in identifying the properties of zoom-whirl orbits.

\begin{figure*}
	\center{
	\subfigure[$q=(1,2,0)$]{\label{hpc9a}
	\includegraphics[width=15cm]{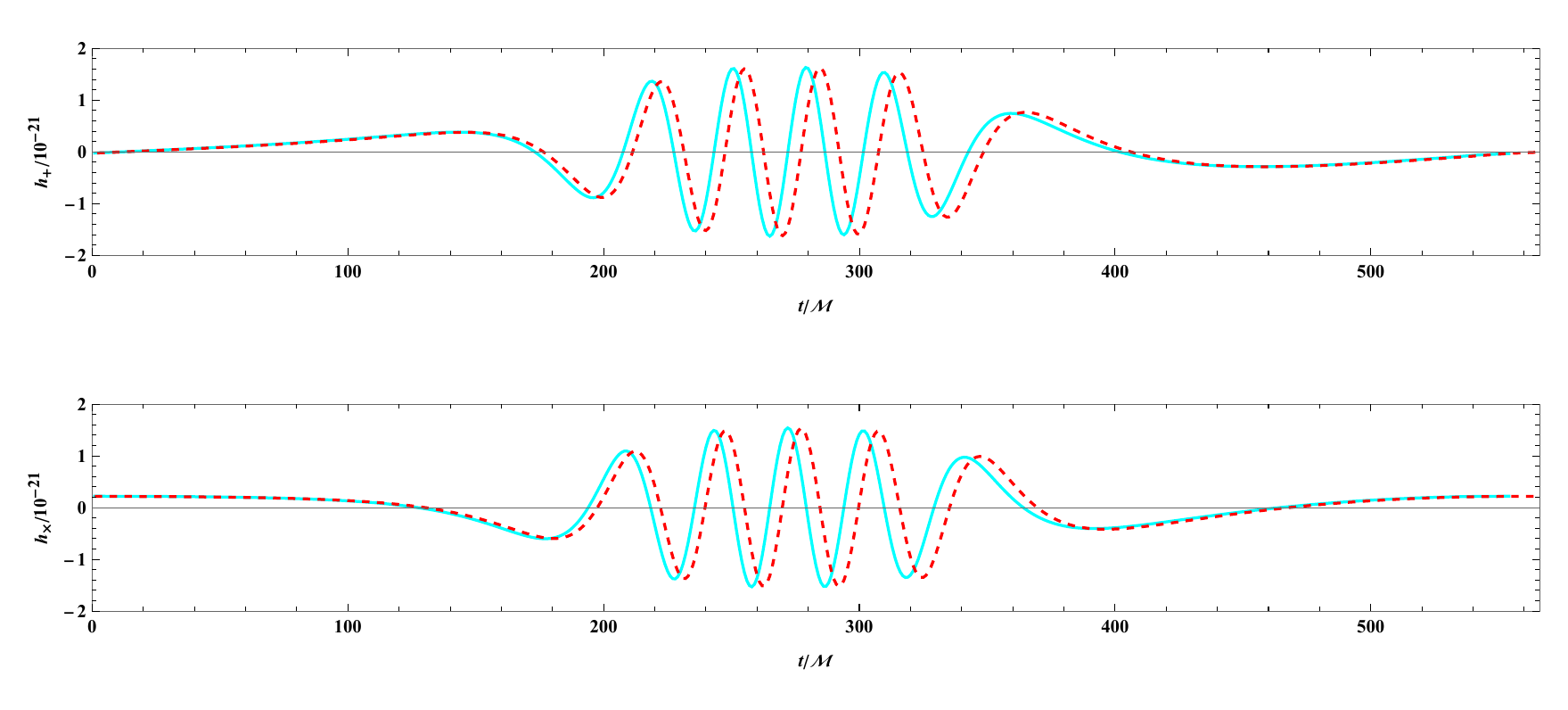}}\\
	\subfigure[$q=(2,1,1)$]{\label{hpc9b}
	\includegraphics[width=15cm]{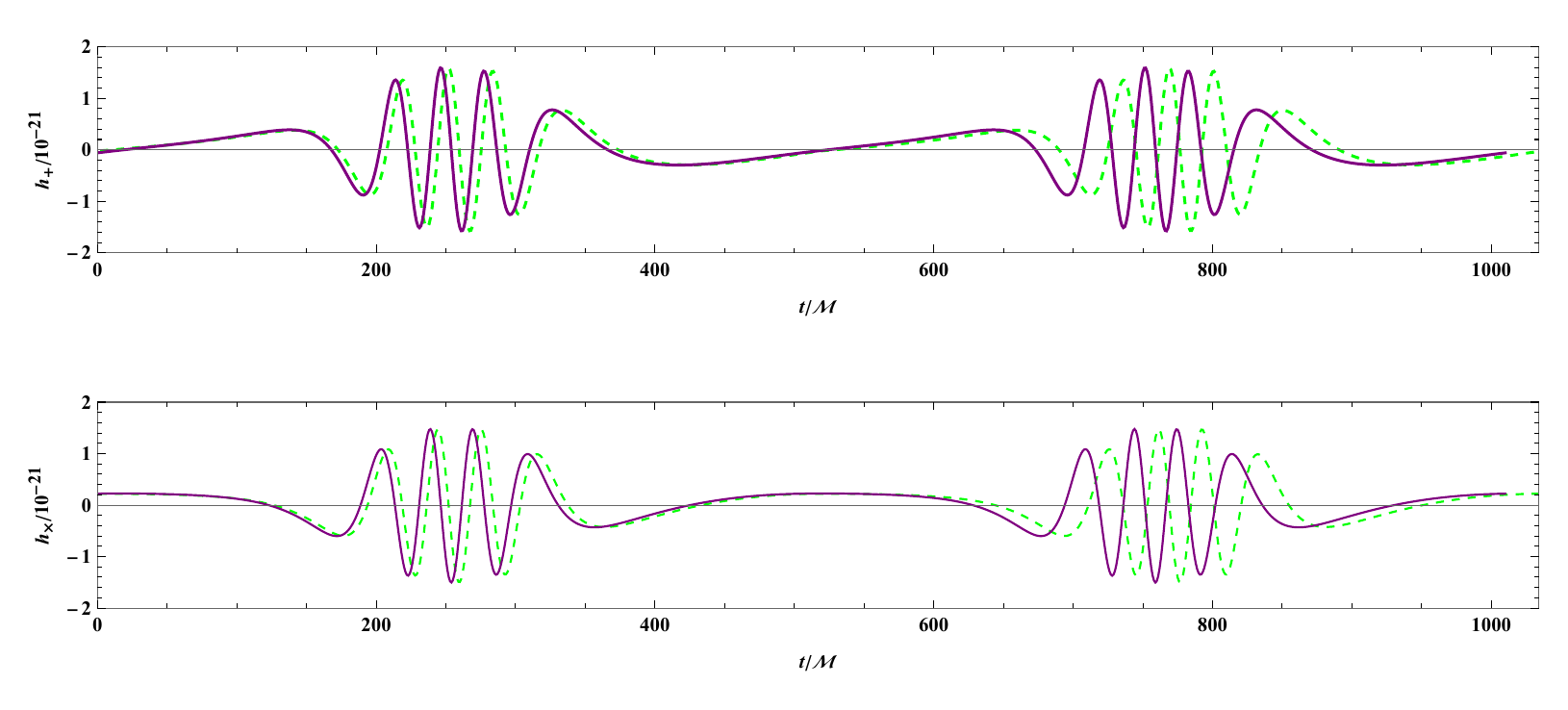}}\\
	\subfigure[$q=(3,2,2)$]{\label{hpc9c}
	\includegraphics[width=15cm]{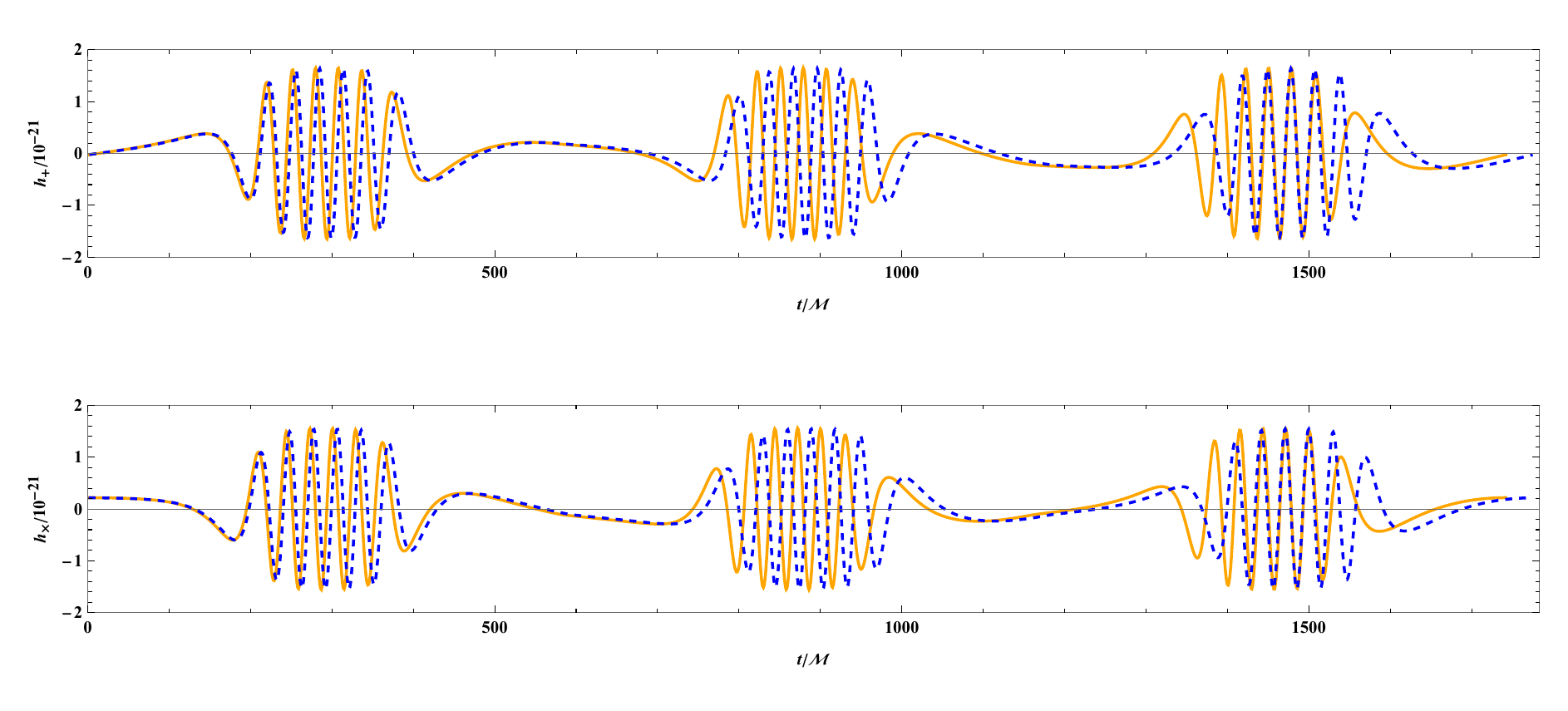}}\\
	}
\captionsetup{justification=raggedright}
\caption{The GW waveforms corresponding to  different hair parameters $\ell$ for different periodic orbit $q=q (z, w, v)$. The solid line in the figure represents $\ell/\mathcal{M}=1$, while the dashed line represents $\ell/\mathcal{M}=1.99$. (a) $q=(1,2,0)$. (b) $q=(2,1,1)$. (c) $q=(3,2,2)$}
\label{fig:waveforms2}
\end{figure*}

Finally, to investigate the detailed impact of the hair parameter on the GW forms associated with different periodic orbits, we compare the waveforms for various $(z, w, v)$ periodic orbits with the hair parameters $\ell/\mathcal{M}= 1$ (solid lines) and $\ell/\mathcal{M}= 1.99$ (dashed lines) in Fig.~\ref{fig:waveforms2}. Some notable features are as follows: The amplitudes of the waveforms for $\ell/\mathcal{M}= 1.99$  and \(\ell/M = 1\) appear similar, indicating that the hair parameter $\ell/\mathcal{M}$ has a negligible effect on the amplitude. However, a noticeable phase shift occurs between the solid and dashed lines, suggesting that $\ell/\mathcal{M}$ affects the phase evolution of the GWs. The oscillatory behavior during the zoom and whirl phases remains consistent between $\ell/\mathcal{M}= 1$ and $\ell/\mathcal{M}= 1.99$, implying that $\ell/\mathcal{M}$ primarily influences the timing rather than the fundamental structure of the waveform. It may be possible to constrain the hair parameter $\ell/\mathcal{M}$ in future high-precision gravitational wave observations through the phase information of periodic orbits. This could be significant for detecting and distinguishing specific physical effects in future GW observations.

\section{Conclusions and discussions}
\label{Conclusion}

In this paper, we explored the potential observational effects of a black hole without the Cauchy horizon, focusing on their periodic orbits and the GWs of EMRIs emitted by the periodic orbits. We began by describing in detail  black holes that are without the Cauchy horizons. Using the Einstein field equation and the gravitational decoupling technique in the Kerr-Schild metric, we separate the mass function in the metric into inner and outer parts, applying appropriate connection conditions to show that a black hole without the Cauchy horizon can be generated by covering a small amount of mass on the event horizon (as long as the hair parameter $\ell$ remains nonzero). Furthermore, when only a small amount of mass is needed on the horizon ($\ell \to 2\mathcal{M}$), the solution approaches the Schwarzschild solution. We then derived the equation of motion from the Lagrangian and obtained the corresponding effective potential for the test particles. By analyzing the effective potential, we examined the properties of the MBO and ISCO. The results show that the orbital radius, energy, and angular momentum of both the MBO and ISCO increase with the hair parameter $\ell$. However, because the black hole external solution asymptotically approaches the Schwarzschild solution at infinity, modified by an exponential factor, the orbital radius, angular momentum, and energy of both the MBO and ISCO rapidly tend toward the Schwarzschild case as $\ell$ increases.

Based on the properties of MBO and ISCO, we investigated the periodic orbits of a black hole without the Cauchy horizon. First, we examined the properties of the rational number $q$ that characterizes the periodic orbit. It is found that $q$ increases with the particle energy and decreases with the particle orbital angular momentum. According to Ref. \cite{Levin:2008mq}, each periodic orbit is described by a set of parameters \((z, w, v)\), and we extended the study to orbits with different hair parameters $\ell$. From Tables. \ref{tab:energylist} and \ref{tab:angmomlisy}, we observed that for periodic orbits with the same $q$, their corresponding energy and orbital angular momentum increase as the hair parameter $\ell$ increases. These results may provide a method to distinguish a black hole free of Cauchy horizon from the Schwarzschild black hole by testing the periodic orbits around the central source.

In addition, we modeled the central supermassive black hole with such a solution and treated the nearby stellar compact objects acting as test particles, forming an EMRIs. The GWs of EMRI radiation are then calculated using the Kludge waveform method based on the particle's trajectory. The $h_{+}$ and $h_{\times}$ waveforms show clear quiet phases during the highly elliptical zoom phases, followed by stronger glitches during the nearly circular whirl phases. These features suggest that, in the future, precise gravitational wave signals covering the full zoom-whirl cycle could aid in identifying the zoom-whirl characteristics of periodic orbits, thereby determining the particle's trajectory. Furthermore, the phase of the GW forms corresponding to periodic orbits can serve as a powerful tool to precisely constrain the parameters of the associated black hole solutions. These properties could be used to identify the orbital structure of EMRI systems and to test/constraint future GW detectors for black holes without Cauchy horizon.


\section*{Acknowledgements}
This work was supported by the National Natural Science Foundation of China (Grants No. 12475055, No. 12075103, No. 12105126, and No. 12247101). Tao Zhu is supported by the National Key Research and Development Program of China under Grant No. 2020YFC2201503, the National Natural Science Foundation of China under Grants No. 12275238 and No. 11675143,  the Zhejiang Provincial Natural Science Foundation of China under Grants No. LR21A050001 and No. LY20A050002, and the Fundamental Research Funds for the Provincial Universities of Zhejiang in China under Grant No. RF-A2019015.

\end{document}